\documentclass{jfm}
\usepackage{graphicx}
\usepackage{epstopdf, epsfig}
\usepackage{amsmath,amssymb} % easier math formulae, align, subequations \ldots
\usepackage{lscape}
\usepackage{subfig}
\usepackage{floatrow}
\usepackage{caption}
\usepackage{enumitem,lipsum}
\usepackage{arydshln}
\usepackage{soul}
\def\IB#1{\boldsymbol{#1}} % use \IB{} for italic bold vector
\def\RE#1{\Rey_{#1}} % macro for Reynolds number
\def\SL#1{\IB{\mathfrak{s}}({\tilde \psi }_{#1})} % psi_{2} contribution to particle velocity
\def\sl#1{\mathfrak{s}_{x}({\tilde \psi }_{#1})} % x-component of above vector

\shorttitle{Inertial migration of an electrophoretic sphere in Poiseuille flow}
\shortauthor{A. Choudhary, T. Renganathan and S. Pushpavanam}

\title{Inertial migration of an electrophoretic rigid sphere in a two-dimensional Poiseuille flow}

\author{A. Choudhary\aff{1},
	T. Renganathan\aff{1},
	\and S. Pushpavanam\aff{1}
	\corresp{\email{spush@iitm.ac.in}}}

\affiliation{\aff{1}Department of Chemical Engineering, Indian Institute of Technology, Chennai,
	
	TN 600036, India}

\begin{document}
	
	\maketitle
	
	\begin{abstract}
		There has been a recent interest in integrating external fields with inertial microfluidic devices to tune particle focusing. In this work, we analyze the inertial migration of an electrophoretic particle in a 2-D Poiseuille flow with an electric field applied parallel to the walls. For a thin electrical double layer, the particle exhibits a slip-driven electrokinetic motion along the direction of the applied electric field, which causes the particle to lead or lag the flow (depending on its surface charge). The fluid disturbance caused by this slip-driven motion is characterized by a rapidly decaying source-dipole field which alters the inertial lift on the particle. We determine this inertial lift using the reciprocal theorem.
		
		Assuming no wall effects, we derive an analytical expression for a ‘phoretic-lift’ which captures the modification to the inertial lift due to electrophoresis. 
%			We also take wall effects into account and find that the analytical expression is valid away from the walls. 
We also take wall effects into account at the leading order, using the method of reflections.
		We find that for a leading particle, the phoretic-lift acts towards the regions of high shear (i.e. walls), while the reverse is true for a lagging particle. 
		Using an order-of-magnitude analysis, we obtain different components of the inertial force and classify them on the basis of the interactions from which they emerge. We show that the dominant contribution to the phoretic-lift originates from the interaction of source-dipole field (generated by the electrokinetic slip at the particle surface) with the stresslet field (generated due to particle's resistance to strain in the background flow).  
		Furthermore, to contrast the slip-driven phenomenon (electrophoresis) from a force-driven phenomenon (buoyancy) in terms of their influence on the inertial migration, we also study a non-neutrally buoyant particle.
		We show that the gravitational effects alter the inertial lift primarily through the interaction of the background shear with the buoyancy induced stokeslet field.
	
	\end{abstract}

	\section {Introduction}
	
	\subsection{Inertial migration}
	Inertial focusing is the cross-stream migration of particles in the presence of finite fluid inertia. This phenomenon was first observed by \citet{segre1961,segre1962a,segre1962b}. Their experiments revealed a cross-stream migration of a dilute suspension of rigid neutrally buoyant particles in a pressure driven flow. The peak concentration of the particles occurred at a radial position of $ \sim $0.6$ \,R_{p} $, where $ R_{p} $ is the pipe radius. \citet{bretherton1962} showed that in the absence of inertia (i.e. Stokes regime), a rigid sphere cannot migrate across streamlines owing to the reversibility of the Stokes flow. Therefore, cross-stream migration was attributed to the fluid inertia.
	
	To qualitatively understand this behavior, the first theoretical studies were conducted for unbounded flows. \citet{rubinow1961} used matched asymptotic expansion to find the inertial force on a particle exhibiting constant rotation in a uniform flow. They found a viscosity independent force in a direction perpendicular to the axis of rotation and the incident flow. In a Poiseuille flow, this force always acts towards the axis of the channel which contradicts the observations of \citet{segre1961}. Therefore their results failed to explain the primary reason behind the lateral migration. Later \citet{saffman1965} derived an expression for the lift on a small sphere exhibiting a relative motion along the streamlines of an unbounded shear flow. Saffman demonstrated that a particle would migrate towards the regions of higher relative velocity; a particle leading (or lagging) in a shear flow would migrate transversely towards the region of low (or high) velocity. Although this study aided in understanding some aspects of the migration observed by Segre-Silberberg (buoyancy and near-wall effects), it could not explain the migration of neutrally buoyant particles.
	
	Migration of a particle in the presence of boundaries was first studied in great generality by \citet{cox1968}. 
	They considered a spherical particle suspended in a three-dimensional Poiseuille flow. They made a crucial observation that, in bounded flows, the viscous stresses are more dominant than the inertial stresses, provided $ \Rey_{V} \ll \kappa  $ ($ \Rey_{V} $ is the particle Reynolds number defined as: $  V' \, a'/{\nu' } $, where $V' $ is the particle velocity relative to the background flow, $ \nu' $ is the kinematic viscosity and $ \kappa $ is the ratio of particle radius $ a' $ to the channel height $ l' $).
	Using a regular perturbation expansion, they derived an expression for migration velocity as a volume integral involving the velocity field in Stokes regime (these velocity fields were represented in terms of Green's function). However, the solution was not provided in an explicit form as it involved numerous integrals. Therefore no conclusions were drawn with respect to lateral migration or equilibrium positions.
	 
	  \citet{ho1974} studied the migration of a neutrally buoyant particle in a 2D Couette and Poiseuille flow and were the first to calculate the lift force explicitly. 
	  Similar to \cite{cox1968} they found that the viscous stresses dominate throughout the channel, provided $\Rey{_p} \ll {\kappa ^2}$. Here $ \Rey{_p} =  U_{max}' \kappa \, a'/{\nu' }  $ is the particle Reynolds number based on the average shear rate ($ U_{max}' $ is the maximum centerline velocity).
	 They employed a regular perturbation expansion in $ \Rey_{p} $ in conjunction with  Lorentz reciprocal theorem to find the inertial lift as a volume integral over creeping flow velocities. These velocities were found using the method of reflections and wall corrections were incorporated using Faxen transformations (originally developed by \citet{faxen1922} for bounded viscous flows). For Poiseuille flow, the equilibrium positions were found to be in agreement with  the observations of \citet{segre1961} i.e. $ \sim $0.6 times the half channel width. Recently, \citet{hood2015} extended the approach of \citet{ho1974} to a 3-D Poiseuille flow. They derived an asymptotic model and calculated lift on the particle as a function of size ratio $\kappa$. Through a numerical analysis, they demonstrated that the methodology developed by Ho and Leal was applicable over a wider range of $\Rey{_p}$ than their theoretical constraint, $ \Rey_{p} \ll \kappa^{2} $. 
	 
	 \citet{vasseur1976} used the general results of \citet{cox1968} to compute the inertial lift on neutrally and non-neutrally buoyant spherical particles in 2D Couette and Poiseuille flows. They approximated the particle as a point force and used an approach based on Green's functions to solve for the flow field. Their results for the migration velocity of neutrally buoyant particles showed a significant deviation from that obtained by \citet{ho1974} near the walls. 

% For bounded flows, the previously discussed theories are applicable for weak inertial effects. For moderate inertia, the walls lie in the Oseen region (where the inertial stresses are comparable to viscous stresses). \citet{vasseur1977lateral} followed the framework of \citet{saffman1965} to determine inertial lift on a sedimenting sphere for $\Rey{_V} \ll 1$ regime. The domain was divided into two regions (viscous dominated and Oseen region) and the method of matched asymptotic expansions was used to obtain the inertial lift. Later \citet{schonberg1989} also followed the approach of \citet{saffman1965} and found the migration of particle in Poiseuille flow. They addressed $ \Rey_{C} \sim O(1) $ regime, where $ \Rey_{C} $ is the channel Reynolds number ($ \Rey_{C}=\Rey_{p} \kappa^{-2} $). They furthermore substantiated an observation made by Segre and Silberberg that, increasing the channel Reynolds number shifts the equilibrium positions towards the walls. 

 \subsection{Electrophoresis tuned inertial focusing}
 In the past decade, numerous experimental studies have focused on the implementation of inertial focusing in microfluidic Lab-on-Chip devices to separate and trap, particles and cells \citep{martel2014inertial,zhang2016}. There has been a recent interest towards the integration of external electric fields with inertial microfluidics to have an active control over particle migration. This can be exploited to separate particles and biological cells on the basis of their ability to acquire a surface charge when suspended in an aqueous (polar) medium. Recent experimental studies by \citet{kim2009,kim2009Three,cevheri2014,yuan2016,xuan2018,rossi2019particle} demonstrated an electric field induced lateral migration by making a polystyrene microsphere lag behind the background Poiseuille flow (this---electric field imposed---relative motion is in addition to the lag induced by the curvature of the background Poiseuille flow and interaction with the boundaries). The particles were observed to focus on the axis (or walls) of the channel when they were made to lag (or lead) the pressure driven flow. 
  
 \floatsetup[figure]{style=plain,subcapbesideposition=top}
 \begin{figure}%
 	\centering
 	\sidesubfloat[]{{\includegraphics[scale=0.22]{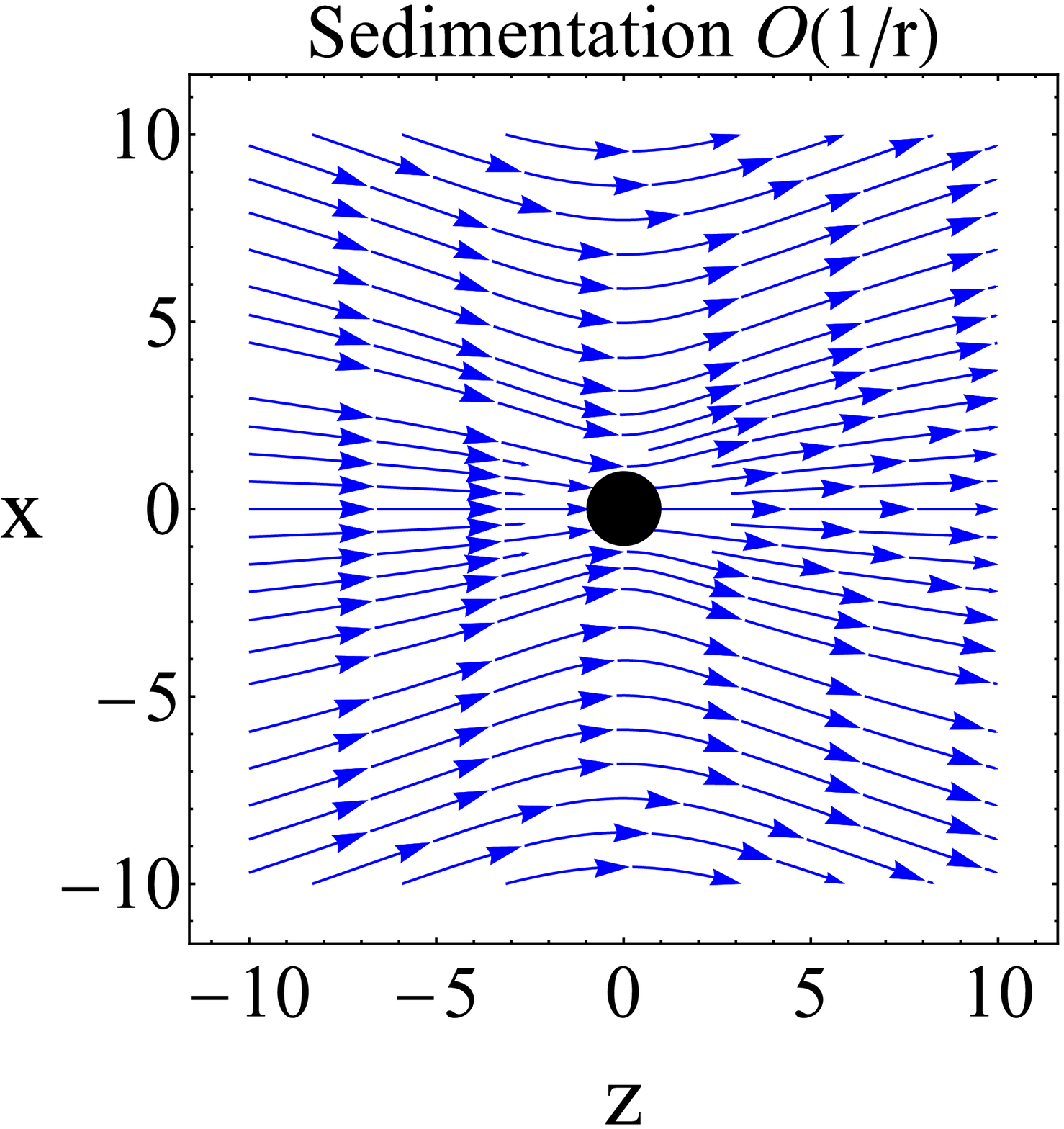} }}%
 	\qquad
 	\sidesubfloat[]{{\includegraphics[scale=0.22]{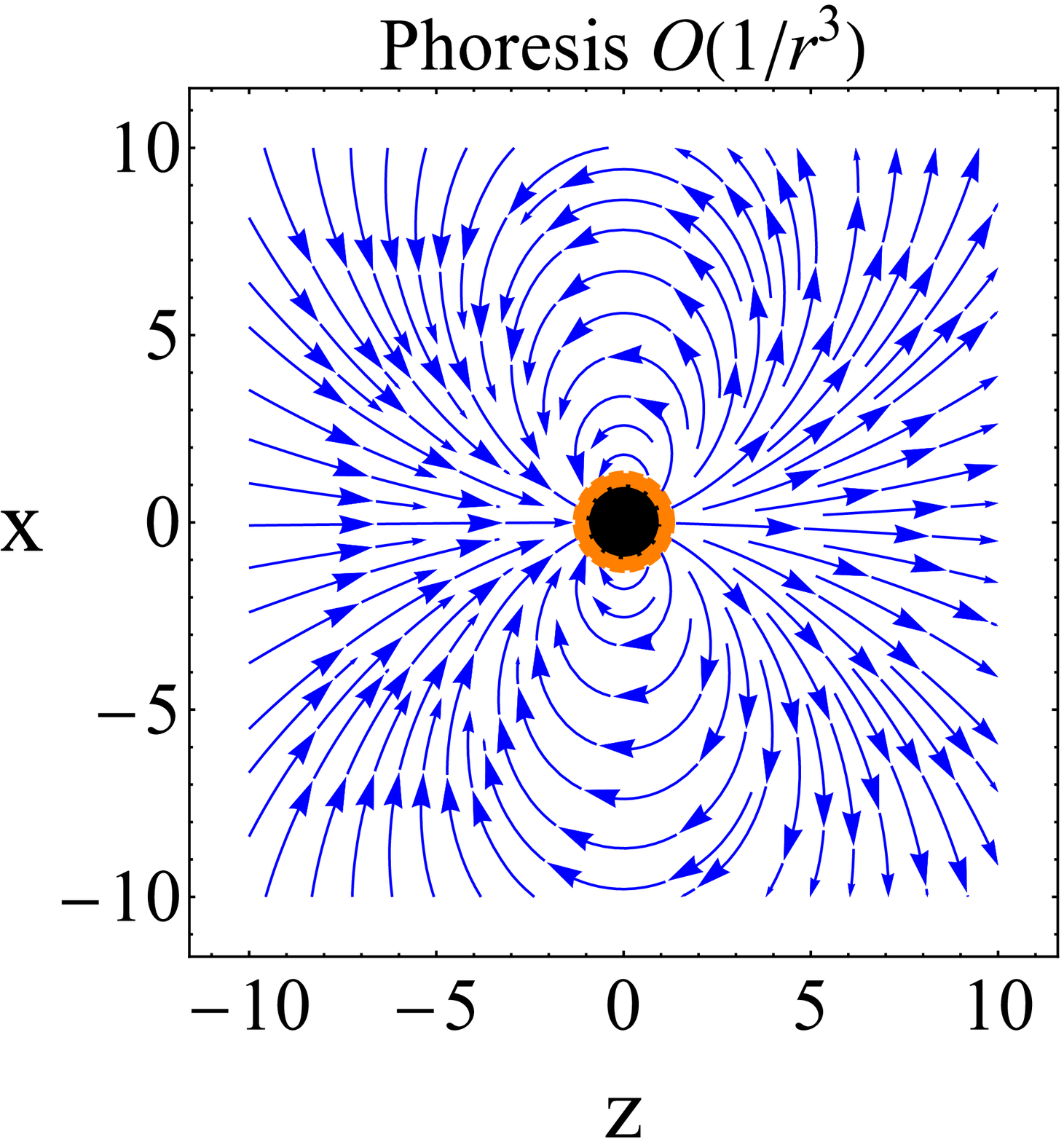} }}%
 	\caption{Disturbance velocity fields around the particle in a quiescent fluid for movements due to (a) forced (sedimentation) and (b) force-free (phoresis) mechanisms. The external field is applied in the horizontal direction.}%
 	\label{fig:1}%
 \end{figure}

A common hypothesis put forth by these experimental studies is that a charged particle under the action of a parallel Coulombic force leads or lags the background flow analogous to a non-neutrally buoyant particle suspended in a Poiseuille flow (with gravity aligned parallel to the flow) which generates a Saffman-lift \citep{saffman1965}. 
 However, a closer examination reveals differences in the two phenomena. The mechanism by which the lag/lead occurs due to gravitational field is fundamentally different from that in an electrolytic medium due to the external electric field. In the latter, a layer of ions (consisting primarily of the counter-ions) forms around the charged particle by electrostatic attraction. This layer is subject to a thermo-electrostatic equilibrium and is known as the electrical double layer. 
 The particles used in the experiments \citep{xuan2018} were a few micrometers in diameter, with zeta potential a few times the thermal potential ($ \sim $25 mV). The double layer ($\lambda_d '$) associated with such particles is a few nanometers thick. Therefore, these systems lie in the thin-double layer limit ($\lambda_d ' / a'\ll 1$).
 In such cases, an external field does not affect the particle directly as a thin ionic layer shields it. The external field however, exerts a Coulomb force on the fluid in the thin layer (as the net charge is non-zero), resulting in a tangential slip on the surface of the particle. As a result, a freely suspended particle moves in the direction opposite to the slip, known as electrophoresis \citep{kirby_2010}.
% The particles used in the experiments \citep{xuan2018} were few micrometers in diameter, with zeta potential a few times the thermal scale ($ \sim $25 mV). The double layer ($\lambda_d '$) associated with such particles is a few nanometers thick. Therefore, these systems lie in thin-double layer ($\lambda_d ' / a'\ll 1$) limit. The screening double layer around the particle renders it neutral to the external field. 

 	\setlength{\dashlinedash}{2pt}
 
 	\begin{table}
 		\renewcommand{\arraystretch}{0.6}
 		 \small
 		\begin{center}
 			\def~{\hphantom{0}}
 			\begin{tabular}{c l l l} \\ \hline \\
 				Investigation                                                     & Description                            & Regime        & Remarks/comments			 \\[10pt] \hline \\

 			\begin{tabular}[t]{c} Rubinow and\\Keller (1961) \end{tabular}       & \begin{tabular}[t]{l}Effect of rotation\\in unbounded uniform\\ flow\end{tabular}      & \begin{tabular}[t]{l} $ \Rey \ll 1 $\end{tabular}     & \begin{tabular}[t]{l} Migration direction perpendicular\\ to incident flow and axis of rotation \end{tabular} \\ \\
 				
 				\begin{tabular}[t]{c} Saffman \\(1965) \end{tabular}       & \begin{tabular}[t]{l} Effect of relative\\velocity in unbounded\\ shear flows \end{tabular}      & \begin{tabular}[t]{c} $ \Rey_{p} \ll 1 $ \end{tabular}     & \begin{tabular}[t]{l} Saffman lift: migration towards\\the regions of higher relative \\velocity \end{tabular} \\ \\
 				
 				\begin{tabular}[t]{c} Ho and Leal\\(1974) \end{tabular}       & \begin{tabular}[t]{l} Neutrally buoyant\\particle in 2D flows \end{tabular}      & \begin{tabular}[t]{c} $ \Rey_{p}\ll \kappa^{2} $\\$ \; \; \kappa \ll 1 $ \end{tabular}     & \begin{tabular}[t]{l} Migration owing to two\\ counteracting forces: \\(i)Wall-lift (towards center)\\ (ii)Shear-gradient-lift (towards walls)\end{tabular} \\ \\
 				
 				\begin{tabular}[t]{c} Vasseur and \\Cox (1976) \end{tabular}       & \begin{tabular}[t]{l} Non-neutrally buoyant\\ particle in 2D flows \end{tabular}      & \begin{tabular}[t]{c} $ \Rey_{V}\ll \kappa $\\$ \; \;  \kappa \ll 1 $ \end{tabular}     & \begin{tabular}[t]{l} Shift in equilibrium position in\\ concordance with \citet{saffman1965} \end{tabular} \\ \\
 				
 				\hdashline \\
 				
 				\begin{tabular}[t]{c} Yariv (2006),\\ (2016) \end{tabular}       & \begin{tabular}[t]{l} Electrophoretic particle\\ near a wall \end{tabular}      & \begin{tabular}[t]{c} \; $ \epsilon \gg 1 $ \end{tabular}     & \begin{tabular}[t]{l} Maxwell stress induced wall\\ repulsion \end{tabular} \\ \\
 				
 				\begin{tabular}[t]{c} Bike and Prieve \\(1995);\\ Schnitzer and \\Yariv (2016) \end{tabular}       & \begin{tabular}[t]{l} Electrophoretic\\particle near a wall,\\suspended in shear flow \end{tabular}      & \begin{tabular}[t]{c} $ P\!e = O(1) $\\ $ \epsilon \ll 1 $ \end{tabular}     & \begin{tabular}[t]{l} Shear-induced electroviscous wall\\ repulsion \end{tabular} \\ 
 				
 				\hdashline \\
 				
 				\begin{tabular}[t]{c} This work \end{tabular}       & \begin{tabular}[t]{l} Neutrally buoyant \\electrophoretic\\particle in plane\\Poiseuille flow \end{tabular}      & \begin{tabular}[t]{c} $ \Rey_{p}\ll \kappa^{2} $\\$ \; \;  \kappa \ll 1 $ \\ $ \; \;\, \epsilon \gg 1 $ \end{tabular}     & \begin{tabular}[t]{l} To potentially explain\\ the observations of \\Kim \textit{et al.} (2009a, 2009b); \\Cevheri and Yoda (2014); \\Yuan \textit{et al}. (2016); \\Li and Xuan (2018); Rossi et al. (2019) \\  \end{tabular}\\ \hline

 			\end{tabular}
 			\caption{A summary of theoretical studies in the past addressing various forces acting on the particle. Table is divided with respect to origin of the lift: (i.) weak inertia and (ii.) electrostatics and electrokinetics. $ \Rey $ is the flow Reynolds number; $ \Rey_{p} $ is the particle Reynolds number based on average shear; $ \Rey_{V} $ is the particle Reynolds number based on particle velocity relative to background flow; $ P\!e $ is the Peclet number defined as $ a'^{2}G'/D_{i}' $ ($ G' $ is the background velocity gradient and $ D_{i}' $ being the ionic diffusion coefficient of $ i $\textsuperscript{th} ionic species); $ \epsilon $ is the particle-wall gap, non-dimensionalized with respect to particle radius $ a' $.}
 			\label{table1}
 		\end{center}
 	\end{table}
 
\normalsize

 In the absence of boundaries, electrophoretic motion corresponding to such particles can be viewed as ‘force-free’ and is fundamentally different from the motion associated with sedimentation \citep{anderson1989}. Figure \ref{fig:1} shows the disturbance a particle creates in a quiescent fluid when subjected to these two mechanisms. The force-free phoretic particle motion occurs due to slipping of counter-ions over its surface. The associated disturbance field is governed by a rapidly decaying source-dipole field ($\sim O(1/r{^3})$). However, the disturbance field generated from the motion induced due to density difference (buoyancy induced motion), on the contrary, is a stokeslet at the leading order (flow field generated due to point force), which decays significantly slower ($\sim O(1/r)$). This stokeslet upon interaction with the background shear gives rise to a buoyancy combined inertial migration \citep{feng1994,matas2004}. Since a direct analogy cannot be extended, the question then arises as to how electrophoresis affects inertial migration and what are the responsible underlying interactions?

 Apart from the inertial lift, there are various other forces which act laterally on an electrophoretic particle. \citet{yariv2006} found that a wall-repulsive electric force acts on the electrophoretic particle when a parallel electric field is applied. 
% This force is due to wall-induced asymmetry in the potential distribution around the particle. 
 As this force arises from the asymmetry in Maxwell stresses, it is decoupled from the hydrodynamics and gets altered in the presence of surface conduction effects \citep{yariv2016}. 
 In addition to this, an externally imposed \textit{mechanical} flow over an electrophoretic particle generates a `streaming potential' (arises due to streaming of ions inside the double layer). 
%  An externally applied pressure field upsets the thermo-electrostatic equilibrium inside the double layer. This externally imposed flow convects the ions in double layer, which consequently generates a surface current leading to a streaming electric field in the bulk.
   \citet{schnitzer2012P2} showed that this consequently generates a streaming electric field which is $ O(\lambda_d '/a')^2 $ for moderate Peclet numbers: $ Pe \sim O(1) $ ($ Pe=\kappa U_{max}'a'/D_{i}' $, where $ D_{i}' $ is the ionic diffusivity). Since we consider double layer to be asymptotically thin ($ \lambda_d '/a' \ll 1 $), the effect of streaming potential is neglected in this work.
   However, in the aforementioned regime, the streaming potential may generate when the particle is very near the walls.
 This modifies the electrokinetic slip, and the associated Newtonian stresses result in a lateral `electro-viscous force' \citep{bike1995,schnitzer2016P3}.
%\citet{schnitzer2016P3} found that this lateral force is dominant in the regions of high shear (i.e. when the particle is in close proximity to the wall).
In this work, we neglect the electro-viscous force and focus on the migration in the bulk of the channel. 
For the same reason, we exclude the forces arising from (i) the van der Waals interactions and (ii) interactions of the double layer between particle and walls.

Table \ref{table1} summarizes various theoretical studies in the past which analyzed lateral migration of neutral and electrophoretic particles in different regimes.
Recent experimental studies \citep{kim2009,kim2009Three,cevheri2014,yuan2016, xuan2018,rossi2019particle} discuss the electric field effect on the migration of electrophoretic particles in Poiseuille flow. 
However, the physical cause and magnitude of this effect remains unclear. 
Also, a closer look at table \ref{table1} sparks a further question: Does a slip-driven motion (electrophoresis) have a different effect on inertial migration than a force-driven motion (movement due to buoyancy)?

 Motivated by the recent experimental studies, the objectives of this work are: (i) to study the inertial lift on an electrophoretic sphere in a combined electro-osmotic Poiseuille flow, (ii) to obtain physical insight on the various interactions which contribute to the migration of an electrophoretic particle, (iii) and to draw comparisons with the buoyancy combined inertial lift.
 The parameters reported by the experimental studies \citep{kim2009,kim2009Three,cevheri2014,yuan2016, xuan2018,yee2018,rossi2019particle} correspond to the regime of weak inertia, where the entire domain is dominated by viscous forces. Therefore, our approach builds on the framework of \citet{ho1974}.

 This work is organized as follows: The problem formulation and different simplifying assumptions are presented in \S \ref{sec:2}. We apply a regular perturbation expansion in $ \Rey_{p} $, and derive the lift force using the Lorentz reciprocal theorem. In \S \ref{sec:3}, the electrostatic potential and velocity fields are found using the method of reflections. We employ Lamb’s general solution \citep{lamb} and Faxen’s transform \citep{faxen1922} to calculate the reflections of fields. In \S \ref{sec:4}, we draw qualitative insights on the migration behaviour by neglecting wall effects and derive an approximate expression for a `phoretic-lift'. In \S \ref{sec:5}, we summarize the solution methodology. We illustrate the migration of an inert particle and validate our results with the previous studies in \S \ref{sec:6}. In \S \ref{sec:7}, the effect of electrophoresis on the inertial lift is analyzed. The different constituents contributing to the lift force are analyzed here. 
 In \S \ref{sec:8}, we study a non-neutrally buoyant particle and draw comparisons with the electrophoretic particle in terms of their influence on the inertial migration.  
 In \S \ref{sec:9}, we discuss a decoupled electrical lift force which stems from the Maxwell stresses \citep{yariv2006} and compare it with the inertial lift. The key results and conclusions are discussed in \S \ref{sec:10}. Appendix and supplementary material provides the details of calculations for the interested reader.

 %%%%%%%%%%%%%%%%%%%%%%%%%%%%%%%%%%%%%%%%%%%%%%%%%%%%%%%%%%%%%%%%%%%%%%%%%%%%%%%%%%%%%%%%
 
 \section{Problem formulation}\label{sec:2}
\begin{figure}
 	\centerline{\includegraphics[scale=0.5]{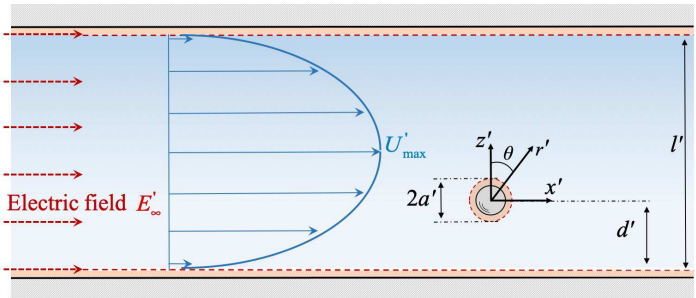}}% Images in 100% size
 	\caption{Schematic depicting the problem configuration and the coordinate system in particle reference frame. The dashed layer around the particle and walls depicts the outer edge of the double layer. Prime denotes the dimensional variables and $\theta$ denotes the polar angle is measured from the $z{'}$-axis. $ a' $ includes the particle radius and thin electrical double layer $ \lambda_d ' $.}
 	\label{fig:2}
 \end{figure}

 The system analyzed in this work is shown in figure \ref{fig:2}. It depicts a neutrally buoyant particle of radius $ a' $ suspended in a combined electro-osmotic Poiseuille flow of an electrolytic solution, moving with a maximum velocity of $U{_{max}'}$ (where $ ' $ denotes dimensional variables). The system is subjected to an externally imposed uniform electric field parallel to the walls ($ \IB{E}_{\infty} ' = E_{\infty} ' {\IB{e}_x}$). Both the particle and wall surface are assumed to have a constant zeta potential. The origin of Cartesian coordinate system is placed at the particle centre, which is located at a distance $d'$ from the bottom wall. The thin double layer is depicted by dashed lines in figure \ref{fig:2}. We analyze the system under the thin-double-layer limit ($ \lambda_d '/a'\ll 1$). The thin double layer assumption implies that the ionic distribution in it is at quasi-equilibrium. Here $\lambda_d '$ denotes the thickness of the electrical double layer. 
 At large zeta potentials, the conduction of ions in the double layer can exceed than that in the bulk. As a result, polarization of ions occurs inside the double layer, formally known as the surface conduction effect \citep{obrien1978,hunter1981}. In this work, since we do not address logarithmically large zeta potentials \citep{schnitzer2012macroscale}, we neglect the surface conduction effects.
%When conduction of ions in the double layer is greater than that in bulk, polarization of ions can occur inside the double layer. This is known as the surface conduction effect \citep{obrien1978,hunter1981}.
% 	In this work, we neglect surface conduction and therefore do not account for `double layer polarization'. This implies the following condition in the thin double layer limit \citep{keh1985}: 
% \begin{equation}
% \left( {{{{\lambda _d'}} \mathord{\left/
% 			{\vphantom {{{\lambda _d '}} a'}} \right.
% 			\kern-\nulldelimiterspace} a'}} \right)\cosh \left( {{{{\cal Z}e'\zeta' } \mathord{\left/
% 			{\vphantom {{{\cal Z}e'\zeta' } {2{k_B}T'}}} \right.
% 			\kern-\nulldelimiterspace} {2{k_B}T'}}} \right) \ll 1.
% \label{2.1}
% \end{equation}
% Here $\cal Z$ denotes the valency of ions in the electrolyte, $e'$ is the electron charge, $\zeta'$ is the particle zeta potential, $ k_{B} $ is the Boltzmann constant, and $ T' $ is the temperature. For monovalent electrolytic solutions (of few millimolars) at room temperature the double layer  typically is a few nanometres thick. For a 10 $\mu$m sized polystyrene particle of 50 mV zeta potential \citep{xuanrevisit} the parameter described in (\ref{2.1}) is 0.002 for $\lambda_d ' \approx 10$nm. This justifies the assumption to neglect surface conduction effects. 

 \subsection{Governing equations}
 We represent the system in non-dimensional variables using $a'$, $\kappa U_{max}'$, $\mu' \kappa U_{max}'/a'$ and $a' |E_{\infty}'|$ as the characteristic scales for length, velocity, pressure and electrostatic potential, respectively.
 Here, $\kappa$ is the size ratio of the particle radius $a'$ to the channel width $l'$, and $\mu'$ denotes the viscosity of the electrolyte. If no external concentration gradient is applied, the electrostatic potential ($ \phi $) in the ‘electro-neutral’ bulk region (zero charge density) is governed by the Laplace equation:
 \begin{equation}
 {\nabla ^2}\phi  = 0.
 \label{A-Poisson}
 \end{equation}
 
% An externally applied pressure field upsets the thermo-electrostatic equilibrium inside the double layer. This externally imposed flow convects the ions in double layer, which consequently generates a surface current leading to a streaming electric field in the bulk. This renders the surface boundary condition for the electrostatic potential, non-homogeneous \citep{yariv2011}. \citet{schnitzer2012P2} showed that this streaming electric field is $ O(\lambda_d '/a')^2 $ for moderate Peclet numbers: $ Pe \sim O(1) $($ Pe=\kappa U_{max}'a'/D_{i}' $, where $ D_{i}' $ is the ionic diffusivity). Since we consider double layer to be asymptotically thin ($ \lambda_d '/a' \ll 1 $) and $ Pe \approx 10^{-1} $ ($ U_{max}' \approx 10^{-3} {\rm{m/s}}, \, \kappa \approx 10^{-2}, \, a' \approx 10^{-5} \, {\rm{m}}, \, D_{i}'  \approx 10^{-9}\rm{m^{2}/s} $ from \cite{xuan2018}), 
Since the effect of streaming potential is neglected in this work, a no-flux condition for the electrostatic potential is imposed at the \textit{surface} of the particle: 
 \begin{equation}
 {{\IB{e}}_r} \bcdot {\bnabla }\phi=0
  \quad \mbox{at\ }\quad r=1.
 \label{A-PotRad}
 \end{equation}
 The electric field remains parallel to the walls (located at $z= s $ and $ z=1-s $, where $ s \equiv d'/l' $), yielding:
 \begin{equation}
{{\IB{e}}_z} \bcdot {\bnabla }\phi=0
\quad \mbox{at walls\ }.
\label{A-PotWall}
\end{equation}
Far away from the particle, the electric potential distribution approaches its undisturbed state ($ \phi_{\infty} \sim -x $):
  \begin{equation}
 \phi  \to {\phi _\infty }
 \quad \mbox{as\ }\quad r\rightarrow \infty.
 \label{A-PotInf}
 \end{equation}\vspace{1mm}

The hydrodynamics of the current problem consists of three time scales: (i) viscous time scale ($ t_{visc}'\sim a'^{2}/\nu' $), (ii) convective time scale of the flow ($ t_{conv}' \sim a'/\kappa U_{max}' $) and (iii) migration (or geometric) time scale ($ t_{mig}' \sim a'/U_{mig}' $).
 Since the time dependence enters the system through the migration time scale, temporal variations can be assumed to be negligible if the migration time scale is much larger than the convective time scale ($ t_{mig}'  \gg t_{conv}' $) \citep{becker1996sedimentation}.
 Since the migration occurs due to weak inertia, the migration velocity is much smaller than the characteristic velocity ($ U_{mig}' \ll \kappa U_{max}' $), which suggests that we can neglect the temporal variations.
 Therefore, the velocity and pressure distribution in the electroneutral bulk are governed by continuity and \textit{quasi-steady} Navier-Stokes equation. The term quasi-steady implies that the dynamics depend only on the instantaneous geometric configuration (i.e. particle position between the walls), provided $ U_{mig}' \ll \kappa U_{max}' $\footnote{A detailed derivation is described in the supplementary information S1.}. 
 In the frame of reference of the moving particle, the flow is governed by:
 \begin{equation}
   \bnabla  \bcdot \IB{U} = 0,\quad {\nabla ^2}\boldsymbol{U} - \bnabla P = {{\mathop{\rm \Rey}\nolimits} _p}\left( { \boldsymbol{U} \bcdot \bnabla \boldsymbol{U}} \right).
 \label{A-Mom}
 \end{equation}
Here, $\Rey_{p}$ is particle Reynolds number defined as: 
 \begin{equation}
\RE{p}=\frac{\rho' U_{\max }'\kappa a'}{\mu'}.
\label{Rey}
\end{equation}

The sphere moves with a---yet to be determined---translational and rotational velocity: $ \IB{U}_{s} $ and $ \IB{\Omega}_{s} $, respectively. Since the domain is infinite in the y-direction, we can consider: $ {\IB{U}_s} = {U_{sx}}{\IB{e}_x} + {U_{sz}}{\IB{e}_z} $  and $ {\IB{\Omega} _s} = {\Omega _{sy}}{\IB{e}_y} $ (i.e. the particle rotates in the direction of the background flow vorticity). 
To account for the electrokinetic effects inside the double layer, we employ a \textit{macroscale} description i.e. the slip-boundary condition derived by \citet{von1903contribution}. 
We obtain the following boundary condition at the particle surface:
 \begin{equation}
\IB{U} = {\IB{\Omega} _s} \times \IB{r} + H\!a{{Z}_p}\IB{\nabla} \phi  
 \quad \mbox{at\ }\quad r=1.
\label{A-VelRad}
\end{equation}
Here, $ Z_{p} $ is the dimensionless particle surface zeta potential ($ \zeta_{p}'/E_{\infty}'a' $). $ H\!a $ is a Hartmann-type number: $ H\!a=\varepsilon' E_{\infty}'^{2} a'/4\pi \mu' \kappa U_{max}' $, which denotes the ratio of electrical energy density to shear stress. Here $\varepsilon'$ is the electrical permittivity. In this work, based on the experimental conditions \citep{kim2009,cevheri2014,xuan2018},we address $ H\!a \sim O(1) $ and $ Z_{p}\sim O(1)  $ regime ($ E_{\infty}' $ $\approx$ $ O(10^{3})-O(10^{4}) $ V/m, $ \zeta_{p}' $= 50 mV, $ a' $=10 $ \mu $m, $ \mu=10^{-3}$ Pa\,s, $ U_{max}'=2.25 $ mm/s, $\kappa=10^{-2}$, $ \varepsilon'=80 \, \varepsilon_{vacuum}' $). \vspace{3mm}

At walls ($ z= s $ and $ z=1-s $), the flow satisfies the Smoluchowski slip condition:
 \begin{equation}
 \IB{U} = H\!a{{Z}_w}\IB{\nabla} \phi -  {\IB{U}_s}
 \quad \mbox{at walls.\ }
\label{A-VelWall}
\end{equation}
 Here, $ Z_{w} $ is the dimensionless wall surface zeta potential ($ \zeta_{w}'/E_{\infty}'a' $). The flow remains undisturbed far away from the particle and satisfies:
 \begin{equation}
\IB{U} \to \IB{V}_{\infty}
\quad \mbox{as\ }\quad r\rightarrow \infty .
\label{A-VelInf}
\end{equation}
 Here, $ \IB{V}_{\infty} $ denotes the undisturbed background flow (Poiseuille flow combined with electro-osmosis) in the frame of reference of the particle:
 \begin{equation}
{\IB{V}_\infty } = \left( {\alpha  + \beta z + \gamma {z^2}} \right){\IB{e}_x} + H\!a{{Z}_w}\IB{\nabla} {\phi _\infty } - {\IB{U}_s}.
\label{A-VInf}
\end{equation}
The constants $ \alpha, \beta$ and $\gamma $ are:
 \begin{equation}
\alpha  = 4{U_{\max }}s\left( {1 - s} \right),{\rm{      }}\beta  = 4{U_{\max }}\kappa \left( {1 - 2s} \right),{\rm{      }}\gamma  =  - 4{U_{\max }}{\kappa ^2},
\label{alpha}
\end{equation}
 where $ \beta $ and $\gamma$ represent the shear and curvature of the background flow.\vspace{2mm}
 
 \textit{Determination of $ \IB{U}_{s} $ and $\IB{\Omega}_{s}$}: The translational and rotational velocities of the sphere are evaluated by acknowledging that the freely suspended \textit{neutral} particle (charged surface plus counter-ions) is force-free and torque-free (see Appendix \ref{appA}.1 for details). Therefore, in the formulation above, the overall (hydrodynamic and electric) force ($\IB{F}_{H}+\IB{F}_{M}$) and torque ($\IB{L}_{H}+\IB{L}_{M}$) on the particle must be zero:
  \begin{equation}
\IB{F} = {\IB{F}_H} + {\IB{F}_M} = \int\limits_{{S_p}} {\IB{n} \bcdot {\IB{\Sigma} _H}\mathrm{d}S}   +  4\pi H\!a\int\limits_{{S_p}} {\IB{n} \bcdot {\IB{\Sigma} _M}\mathrm{d}S}  = \IB{0},
 \label{A-F}
 \end{equation}
 \begin{equation}
\IB{L} = {\IB{L}_H} + {\IB{L}_M} = \int\limits_{{S_p}} {\IB{n} \times \left( {\IB{n} \bcdot {\IB{\Sigma} _H}} \right)\mathrm{d}S}   +  4\pi H\!a\int\limits_{{S_p}} {\IB{n} \times \left( {\IB{n} \bcdot {\IB{\Sigma} _M}} \right)\mathrm{d}S}  = \IB{0}.
 \label{A-L}
 \end{equation}
 The above force and torque have been normalized with $ \mu' a' \kappa U_{max}' $ and $ \mu' a'^{2} \kappa U_{max}' $, respectively. Here, $ \IB{\Sigma}_{H} $ is the dimensionless Newtonian stress tensor which contributes to hydrodynamic force; $ \IB{\Sigma}_{M} $ is the dimensionless Maxwell stress tensor which contributes to the electric force:
  \begin{equation}
{\IB{\Sigma} _H} =  - P\mathsfbi{I} + \IB{\nabla} \IB{U} + {\left( {\IB{\nabla} \IB{U}} \right)^{T}} \; \mbox{and\ } {\IB{\Sigma} _M} = \IB{\nabla} \phi  \IB{\nabla} \phi  - \frac{1}{2}\left( {\IB{\nabla} \phi  \bcdot \IB{\nabla} \phi } \right)\mathsfbi{I}.
 \label{A-Stress}
 \end{equation}
% and $ \IB{\Sigma}_{M} $ is the dimensionless Maxwell stress tensor which contributes to the electric force:
% \begin{equation}
% {\IB{\Sigma} _M} = \IB{\nabla} \phi  \IB{\nabla} \phi  - \frac{1}{2}\left( {\IB{\nabla} \phi  \bcdot \IB{\nabla} \phi } \right)\mathsfbi{I}.
% \label{A-MaxStress}
% \end{equation}
 Here, $ \mathsfbi I $ is the second order identity tensor and the superscript $ T $ denotes transpose.\\

 \subsection{Equations governing the disturbance} 
 \vspace{2mm}
 The solution to equations (\ref{A-Poisson})-(\ref{A-Mom}), (\ref{A-VelRad})-(\ref{A-VelInf}), (\ref{A-F}) and (\ref{A-L}) is sought in terms of disturbance variables: electrostatic potential ($ \psi=\phi-\phi_{\infty} $), velocity ($ \IB{v}=\IB{U}-\IB{V}_{\infty} $) and pressure ($ p=P-P_{\infty} $). These represent the deviation of actual fields from the undisturbed fields. 
The disturbance electrostatic potential is governed by:
\begin{equation}
{\IB{\nabla} ^2}\psi  = 0,
\label{D-Poisson}
\end{equation}
and is subject to the following boundary conditions:
\begin{equation}
{\IB{e}_r} \bcdot \IB{\nabla} \psi  =  - {\IB{e}_r} \bcdot \IB{\nabla} {\phi _\infty }
\quad \mbox{at\ } \quad r=1,
\label{D-PotRad}
\end{equation}
\begin{equation}
{\IB{e}_z} \bcdot \IB{\nabla} \psi  =0
\quad \mbox{at walls\ },
\label{D-PotWall}
\end{equation}
\begin{equation}
\psi \rightarrow 0
\quad \mbox{as\ } r\rightarrow \infty.
\label{D-PotInf}
\end{equation}
The disturbance velocity is governed by:
\begin{equation}
\bnabla  \bcdot \IB{v} = 0,\quad {\nabla ^2}\IB{v} - \bnabla p = \Rey_{p}\left( { \IB{v} \bcdot \bnabla \IB{v} + \IB{v} \bcdot \bnabla {\IB{V}_\infty } + {\IB{V}_\infty } \bcdot \bnabla \IB{v}} \right),
\label{D-Mom}
\end{equation}
 and is subject to the following boundary conditions:
 \begin{equation}
 \IB{v} = {\IB{\Omega}_s} \times \IB{r} + H\!a{{Z}_p}\bnabla \left( {\psi  + {\phi _\infty }} \right) - {\IB{V}_\infty }
 \quad \mbox{at\ } \quad r=1,
 \label{D-VelRad}
 \end{equation}
 \begin{equation}
 \IB{v} = H\!a{{Z}_w}\bnabla \psi 
 \quad \mbox{at walls,\ }
 \label{D-VelWall}
 \end{equation}
 \begin{equation}
 \IB{v} \rightarrow \IB{0}
 \quad \mbox{as\ } r\rightarrow \infty.
 \label{D-VelInf}
  \end{equation}
 The disturbance hydrodynamic and Maxwell  stresses are defined as:
 \begin{equation}
 {\IB{\sigma} _H} =  - p\mathsfbi I + \bnabla \IB{v} + {\left( {\bnabla \IB{v}} \right)^T} 
 \quad {\IB{\sigma} _M} = \bnabla \psi \bnabla \psi  - \frac{1}{2}\left( {\bnabla \psi  \bcdot \bnabla \psi } \right)\mathsfbi I.
 \label{D-Stress}
 \end{equation}
 Since only the disturbance stresses must contribute to force and torque on the particle, we write:
 \begin{equation}
 \IB{F} = {\IB{F}_H} + {\IB{F}_M} = \int\limits_{{S_p}} {\IB{n} \bcdot {\IB{\sigma} _H}\mathrm{d}S}   +  4\pi H\!a\int\limits_{{S_p}} {\IB{n} \bcdot {\IB{\sigma} _M}\mathrm{d}S}  = \IB{0},
 \label{D-F}
 \end{equation}
 \begin{equation}
 \IB{L} = {\IB{L}_H} + {\IB{L}_M} = \int\limits_{{S_p}} {\IB{n} \times \left( {\IB{n} \bcdot {\IB{\sigma} _H}} \right)\mathrm{d}S}  + 4\pi H\!a\int\limits_{{S_p}} {\IB{n} \times \left( {\IB{n} \bcdot {\IB{\sigma} _M}} \right)\mathrm{d}S}  = \IB{0}.
 \label{D-L}
 \end{equation}

\subsection{Regular perturbation expansion}
The primary objective of this work is to obtain the lift associated with inertia for an asymptotically small $ \Rey_{p} $. Towards this, we perform a regular perturbation expansion in $ \Rey_{p} $. This expansion is valid only when the inertia persists as a perturbation throughout the domain. Therefore, the wall particle distance must be shorter than the length scale where the inertial force balances viscous force \citep{guazzelli2011}.
The order of magnitude of the ratio of inertial to viscous forces shows that the length scale associated with the bulk shear ($ a'/\sqrt{\Rey_{p}}$) is much shorter than that based on the electrokinetic slip ($ a'/H\!a Z_{p} \Rey_{p} $). Therefore, based on the inertial length scale of bulk shear, we arrive at the following constraint for the entire domain to be dominated by viscous forces:
 \begin{equation}
O(\Rey_{p}) \ll \kappa^{2}.
\label{Constraint}
\end{equation}
 This allows us to seek the solution for disturbance velocity, pressure, translational and rotational velocities as a perturbation expansion in $ \Rey_{p} $:
 \begin{equation}
 \begin{array}{l}
 \IB{v} = {\IB{v}^{(0)}} + \Rey_{p}{\IB{v}^{(1)}} +\dotsm\qquad \quad p = {p^{(0)}} + \Rey_{p}{p^{(1)}} + \dotsm\\
 {\IB{U}_s} = \IB{U}_s^{(0)} + \Rey_{p}\IB{U}_s^{(1)} +\dotsm\quad {\IB{\Omega} _s} = \IB{\Omega} _s^{(0)} + \Rey_{p}\IB{\Omega} _s^{(1)} + \dotsm.
 \end{array}
 \label{Pert}
 \end{equation}
 We substitute the above expansion into the equations governing hydrodynamics (\ref{D-Mom})-(\ref{D-VelInf}). We obtain the $ O(1) $ problem as:
 \begin{equation}
 \left. \begin{array}{l}
\qquad \; \quad \bnabla  \bcdot {\IB{v}^{(0)}} = 0, \\
{\nabla ^2}{\IB{v}^{(0)}} - \bnabla {p^{(0)}} = \IB{0},\\
\qquad \qquad \quad{\IB{v}^{(0)}} = \IB{\Omega} _s^{(0)} \times \IB{r} + H\!a{Z_p}\bnabla \left( {\psi  + {\phi _\infty }} \right) - \IB{V}_\infty ^{(0)} \quad \mbox{at\ } r = 1,\\
\qquad \qquad \quad{\IB{v}^{(0)}} = H\!a{Z_w}\bnabla \psi \quad \mbox{at walls},\\
\qquad \qquad \quad{\IB{v}^{(0)}} \to \IB{0} \quad \mbox{as\ } r \to \infty,
 \end{array} \right\}
 \label{Order0}
 \end{equation}
 and the $ O(\Rey_{p}) $ problem as:
  \begin{equation}
 \left. \begin{array}{l}
 \qquad \; \quad \bnabla  \bcdot {\IB{v}^{(1)}} = 0, \\
  {\nabla ^2}{\IB{v}^{(1)}} - \bnabla {p^{(1)}} =  {\IB{v}^{(0)}} \bcdot \bnabla {\IB{v}^{(0)}} + \IB{V}_\infty ^{(0)} \bcdot \bnabla {\IB{v}^{(0)}} + {\IB{v}^{(0)}} \bcdot \bnabla \IB{V}_\infty ^{(0)},\\
\qquad \qquad \quad {\IB{v}^{(1)}} = \IB{U}_{s}^{(1)} + \IB{\Omega} _s^{(1)} \times \IB{r}\quad \mbox{at\ } r = 1,\\
 \qquad \qquad \quad{\IB{v}^{(1)}} = \IB{0} \quad \mbox{at walls},\\
 \qquad \qquad \quad{\IB{v}^{(1)}} \to \IB{0} \quad \mbox{as\ } r \to \infty.
 \end{array} \right\}
 \label{Order1}
 \end{equation}
 We do not perform a similar expansion for the electrostatic potential because the problem governing electrostatics is decoupled from the flow field. \vspace{1mm}
 
 Owing to (\ref{Pert}), the hydrodynamic force and torque can be expanded as:
\begin{equation}
{\IB{F}_H} = \IB{F}_H^{\left( 0 \right)} + \Rey_{p}\IB{F}_H^{\left( 1 \right)} \quad \mbox{and\ } {\IB{L}_H} = \IB{L}_H^{\left( 0 \right)} + \Rey_{p}\IB{L}_H^{\left( 1 \right)}.
\label{HydFL}
\end{equation} 
Here, $ \IB{F}_{H}^{(0)} $ represents the hydrodynamic force in the Stokes regime and therefore acts in the $ x $-direction alone.
 As the flow is two-dimensional, the hydrodynamic force due to inertia ($ \IB{F}_{H}^{(1)} $) is of the following form: $ \{ F_x^{\left( 1 \right)}, 0, F_z^{\left( 1 \right)}\} $. The cross-stream inertial migration arises from $ F_{z}^{(1)} $, whose evaluation is the primary objective of this work.
 One way to obtain this is by solving (\ref{Order0}) and (\ref{Order1}), and evaluating $ \Rey_{p} \, \IB{e}_{z} \bcdot \int\limits_{{S_p}} {\IB{n} \bcdot \IB{\sigma} _H^{(1)}\mathrm{d}S} $. Alternatively, an application of the Lorentz reciprocal theorem allows us to determine the inertial lift force ($ F_{I} \equiv \Rey_{p} F_{z}^{(1)} $)  associated with $ \IB{v}^{(1)} $, without solving for (\ref{Order1}). In the next section, we will elaborate on this approach which was proposed originally by \citet{ho1974}.

 \subsection{Lift force}
 
 The reciprocal theorem relates the properties of an unknown Stokes flow ($ \IB{\sigma},\IB{v} $) to a known test flow field ($ \IB{\sigma}^{t},\IB{u}^{t}  $), provided both fields correspond to the same geometry. The test field is taken to be that generated by a sphere moving in the positive z-direction with unit velocity $ \IB{e}_{z} $ in a quiescent fluid between the walls. The test field is governed by:
 \begin{equation}
 \left. \begin{array}{l}
 \bnabla  \bcdot {\IB{u}^t} = 0, \; \; \; {\nabla ^2}{\IB{u}^t} - \bnabla {p^t} = \IB{0},\\
 \; \; \,\quad {\IB{u}^t} = {\IB{e}_z} \quad \mbox{at\ } r = 1,\\
 \; \; \,\quad{\IB{u}^t} = \IB{0} \quad \; \, \mbox{at the walls,\ }\\
 \; \; \, \quad{\IB{u}^t} \to \IB{0 }\quad  \;  \mbox{at r} \to \infty .
 \end{array} \right\}
 \label{Test}
 \end{equation} 
 The generalized reciprocal theorem  \citep{kim2013} states:
 \begin{equation}
 \int_{{S_p}} {\IB{v} \bcdot \left( {{\IB{\sigma} ^t} \bcdot {\IB{n}^t}} \right)\mathrm{d}S}  + \int_{{V_f}} {\IB{v} \cdot \left( {\bnabla  \bcdot {\IB{\sigma} ^t}} \right)\mathrm{d}V}  = \int_{{S_p}} {{\IB{u}^t} \bcdot \left( {\IB{\sigma}  \bcdot \IB{n}} \right)\mathrm{d}S}  + \int_{{V_f}} {{\IB{u}^t} \bcdot \left( {\bnabla  \bcdot \IB{\sigma} } \right)\mathrm{d}V}.
 \label{Reci}
 \end{equation} 
 Here $ \IB{\sigma}^{t} $ and $\IB{\sigma}$ are the stress tensors corresponding to the test Stokes field and unknown Stokes field. 
 The Navier-Stokes equations (\ref{Order1}) and (\ref{Test}) are expressed in terms of stress tensors as: 
 \begin{equation}
 \bnabla  \bcdot {\IB{\sigma}_{H} ^{(1)}} = {\IB{v}^{(0)}} \bcdot \bnabla {\IB{v}^{(0)}} + \IB{V}_\infty ^{(0)} \bcdot \bnabla {\IB{v}^{(0)}} + {\IB{v}^{(0)}} \bcdot \bnabla \IB{V}_\infty ^{(0)},
\mbox{ and \ } \bnabla  \bcdot {\IB{\sigma} ^t} = \IB{0}.
 \label{dump}
 \end{equation} 
 Substituting $ \IB{\sigma}_{H}^{(1)} $ as $ \IB{\sigma} $ in (\ref{Reci}) and using the boundary conditions expressed in (\ref{Order1}), we obtain:
 \begin{equation}
\begin{array}{l}
{\IB{U}_s}^{(1)} \bcdot \int_{{S_p}} {\left( {{\IB{\sigma} ^t} \bcdot {\IB{n}^t}} \right)\mathrm{d}S}  +  \int_{{S_p}} ({\IB{\Omega} _s}^{(1)} \times \IB{r}) \bcdot ({\IB{\sigma} ^t} \bcdot {\IB{n}^t}) \\
\\
= {\IB{e}_z} \bcdot \int_{{S_p}} {\left( {{\IB{\sigma} _{H}^{(1)}} \bcdot \IB{n}} \right)\mathrm{d}S}  + \int_{{V_f}} {{\IB{u}^t} \bcdot \left( {{\IB{v}^{(0)}} \bcdot \bnabla {\IB{v}^{(0)}} + {\IB{V}^{(0)}_\infty } \bcdot \bnabla {\IB{v}^{(0)}} + {\IB{v}^{(0)}} \bcdot \bnabla {\IB{V}^{(0)}_\infty }} \right)\mathrm{d}V}.
\end{array}
 \label{SimpleReci}
 \end{equation} 
The surface integral in the first term on the left-hand side denotes the hydrodynamic force on the sphere due to the z-direction motion $ \approx  - 6\pi ({1 + O(\kappa )}){\IB{e}_z}$ (with first order wall correction). The second term can be shown to be zero by invoking symmetry arguments. Also, since the particle is freely suspended and neutrally buoyant, the first term on the right-hand side is zero. This yields the following expression for the migration velocity of the particle due to inertia:
 \begin{equation}
 U_{s z}^{\left( 1 \right)} = \frac{{ - 1}}{{6\pi \left( {1 + O(\kappa )} \right)}}\int_{{V_f}} {{\IB{u}^t} \bcdot \left( {{\IB{v}^{(0)}} \bcdot \bnabla {\IB{v}^{(0)}} + {\IB{V}^{(0)}_\infty } \bcdot \bnabla {\IB{v}^{(0)}} + {\IB{v}^{(0)}} \bcdot \bnabla {\IB{V}^{(0)}_\infty }} \right)\mathrm{d}V}.
 \label{MigVel}
 \end{equation}
 Following \citet{ho1974}, at \textit{the present order of approximation} (for migration of $ O(\Rey_{p}) $), we alternatively prescribe $ U_{sz}\equiv 0 $ in (\ref{SimpleReci}). In other words, the particle is not allowed to migrate across streamlines \footnote{Since the particle is free to move in the flow direction, at $ O(\Rey_{p}^{0})  $ the problem remains force-free in $ x $-direction and torque free in $ y $-direction \citep{ho1974}.}. This implies that a finite inertial force $ \IB{F}_{I} $ acts on the particle:
\begin{equation}
\begin{array}{l}
{F_I} \equiv \Rey_{p}\left( {\IB{e}_z \bcdot \int_{{S_p}} {\left( {{\IB{\sigma}_{H}^{(1)}} \bcdot \IB{n}} \right)\mathrm{d}S} } \right)\\
\quad \; \,=  - \Rey_{p}\int_{{V_f}} {{\IB{u}^t} \bcdot \left( {{\IB{v}^{(0)}} \bcdot \bnabla {\IB{v}^{(0)}} + \IB{V}_\infty ^{(0)} \bcdot \bnabla {\IB{v}^{(0)}} + {\IB{v}^{(0)}} \bcdot \bnabla \IB{V}_\infty ^{(0)}} \right)\mathrm{d}V} .
\end{array}
\label{MigF}
\end{equation}
 
 We have now obtained the inertia induced lift force as a volume integral which can be evaluated by solving (\ref{Order0}) (equations governing the creeping flow $ \IB{v}^{(0)} $), (\ref{D-Poisson})-(\ref{D-PotInf}) (equations governing the electrostatic potential $ \psi $) and (\ref{Test}) (equations governing the  test field). In the next section, we evaluate the disturbance fields and the test field.

  %%%%%%%%%%%%%%%%%%%%%%%%%%%%%%%%%%%%%%%%%%%%%%%%%%%%%%%%%%%%%%%%%%%%%%%%%%%%%%%%%%%%%%%%
 \section{Evaluation of disturbances}\label{sec:3}
 In this work, the disturbance potential ($ \psi $) is independent of the hydrodynamics. It affects the hydrodynamics through the slip conditions at the particle and wall surfaces (see eq. \ref{Order0}). Therefore, the electrostatic potential is evaluated first.

 \subsection{Electrostatic potential}
 When the particle is not too close to the wall ($ \kappa/s \ll 1 $), the solution can be sought in terms of successive reflections \citep{brenner1962}. The first reflection represents the disturbance due to a particle in an unbounded domain. The second reflection corrects the previous disturbance by imposing the boundary condition at the walls and ignoring the particle. Subsequently, each reflection imposes the boundary condition alternatively at the particle and walls. Every successive pair of reflections increases the accuracy by $ O(\kappa) $ \citep{happel2012low}. This iterative process is performed until a desired accuracy is attained. We seek the disturbance potential as a sum of reflections:
 \begin{equation}
  \psi  = {\psi _1} + {\psi _2} + \dotsm.
 \label{MorPot}
 \end{equation}
 Here, $ \psi_{i} $ represents the i\textsuperscript{th} reflection. Substituting this in (\ref{D-Poisson})-(\ref{D-PotInf}), we obtain for the first reflection:
 \begin{equation}
 \left. \begin{array}{l}
\; \, \quad {\nabla ^2}{\psi _1} = 0,\\
 {\IB{e}_r} \bcdot \bnabla {\psi _1} =  - {\IB{e}_r} \bcdot \bnabla {\phi _\infty }\quad \mbox{at\ } r=1,\\
\; \;\, \qquad {\psi _1} \to 0 \quad \mbox{as\ } r \to \infty,
 \end{array} \right\}
 \label{Pot1}
 \end{equation}
 and for the second reflection:
 \begin{equation}
 \left. \begin{array}{l}
\; \,\quad {\nabla ^2}{\psi _2} = 0,\\
 {\IB{e}_z} \bcdot \bnabla {\psi _2} =  - {\IB{e}_z} \bcdot \bnabla \psi_{1} \quad \mbox{at walls.\ }
 \end{array} \right\}
 \label{Pot2}
 \end{equation}
 
\textit{ Solution to} $ \psi_{1} $: Equation (\ref{Pot1}) suggests that $ \psi_{1} $ is a harmonic function which decays as $ r \to \infty $. The boundary condition suggests that it must be linear in the ‘driving force’: $ \bnabla {\phi _\infty } (=-{\IB{e}_x}) $. Therefore, the solution is sought in terms of spherical solid harmonics \citep{guazzelli2011} as:
 \begin{equation}
 {\psi _1} = \frac{1}{2}\frac{\IB{r}}{{{r^3}}}.\bnabla {\phi _\infty } =  - \frac{1}{2}\frac{x}{{{r^3}}}.
 \label{PotSol1}
 \end{equation}
 
\textit{ Solution to} $ \psi_{2} $: To evaluate $ \psi_{2} $, we adopt the approach devised by \citet{faxen1922} in the context of bounded viscous flows. Using this, the disturbance around the particle is expressed in an integral form which satisfies the boundary conditions at the walls. The first reflection is characterized by particle scale ($ a' $). The second reflection $ \psi_{2} $ is characterized by a length scale ($ l' $) which is $ O(1/\kappa) $, because the walls are remotely located. Therefore, the coordinates for second reflection are stretched, and are termed as `outer' coordinates. These outer coordinates (denoted by capital letters) are defined as:
 \begin{equation}
 R = r\kappa \quad X = x\kappa \quad Y = y\kappa \quad Z = z\kappa.
 \label{Rescale}
 \end{equation}
Using Faxen's transformations (see Appendix \ref{appA}.2), we obtain the second reflection:
 \begin{equation}
 {\tilde \psi _2} = \frac{{{\kappa ^2}}}{{2\pi }}\int\limits_{ - \infty }^{ + \infty } {\int\limits_{ - \infty }^{ + \infty } {{\mathrm{e}^{\mathrm{i}\Theta }}\left( {{\mathrm{e}^{\left( { - \frac{{\lambda Z}}{2}} \right)}}{b_2} + {\mathrm{e}^{\left( { + \frac{{\lambda Z}}{2}} \right)}}{b_3}} \right)\left( {\frac{{\mathrm{i}\xi }}{\lambda }} \right) \mathrm{d}\xi \mathrm{d}\eta } }.
 \label{PotSol2}
 \end{equation}
 Here $ b_{1} $ and $ b_{2} $ are:
 \begin{equation}
 {b_2} = \frac{1}{8}\frac{{\left( {1 + {\mathrm{e}^{\left( {1 - s} \right)\lambda }}} \right)}}{{ - 1 + {\mathrm{e}^\lambda }}},\quad {b_3} = \frac{1}{8}\frac{{\left( {1 + {\mathrm{e}^{s\lambda }}} \right)}}{{ - 1 + {\mathrm{e}^\lambda }}}.
 \label{b1b2}
 \end{equation}
 
 The next reflection ($ \psi_{3} $) is $ O(\kappa^{3}) $. As we restrict $ \kappa \ll 1 $, the first two reflections capture the leading order contribution to the disturbance potential.\vspace{2mm}

\textit{Solution of }$ \IB{F}_{M} $ \textit{and} $ \IB{L}_{M} $: Before determining the inertial lift $ F_{I} $ through the evaluation of velocity disturbances in the creeping flow limit, we find the electrical force and torque on the particle caused by Maxwell stress. These are used later in the calculation of $ \IB{U}_{s}^{(0)} $ and $ \IB{\Omega}_{s}^{(0)} $ (see \S\ref{sec:3}.3).

We substitute (\ref{PotSol1}) and (\ref{PotSol2}) in both (\ref{D-F}) and (\ref{D-L}). Upon simplification (see Appendix \ref{appA}.3) we obtain:
 \begin{equation}
 {\IB{F}_M} = 4\pi H\!a\left( {\frac{{3\pi }}{{16}}{\kappa ^4}\left(\mathfrak{Z}{\left({4,s} \right) - \mathfrak{Z} \left( {4,1 - s} \right)} \right)} \right) \IB{e}_{z}  +  O\left( {{\kappa ^7}} \right),
 \label{FM}
 \end{equation}
 \begin{equation}
 {\IB{L}_M} = \IB{0}.
 \label{LM}
 \end{equation}
 Here, $ \mathfrak{Z} $ is the generalized Riemann zeta function \citep{abramowitz1972}. This leading order electrical force acts only in the z-direction and always away from the walls (i.e. positive below the channel centerline and vice-versa). As the particle approaches walls ($ s \to 0 \mbox{ or\ } 1 $), (\ref{FM}) asymptotically matches with the single wall expression derived by \citet{yariv2006}. We also find that the Maxwell stresses exert no torque on the particle, at the present level of approximation.
 
 Having obtained the leading order potential distribution, we next evaluate the velocity disturbance.

 \subsection{Velocity disturbance}
 We seek the disturbance ($ \IB{v}^{(0)}, p^{(0)} $) as successive reflections:
 \begin{equation}
 {\IB{v}^{\left( 0 \right)}} = \IB{v}_1^{\left( 0 \right)} + \IB{v}_2^{\left( 0 \right)} + \IB{v}_3^{\left( 0 \right)} +  \cdots, \quad 
 {p^{\left( 0 \right)}} = p_1^{\left( 0 \right)} + p_2^{\left( 0 \right)} + p_3^{\left( 0 \right)} +  \cdots.
 \label{MorVel}
 \end{equation}
Upon substituting the above expansion in (\ref{Order0}), we obtain the following set of problems:
\begin{equation}
\left. \begin{array}{l}
\bnabla  \bcdot \IB{v}_1^{\left( 0 \right)} = 0, \quad {\nabla ^2}\IB{v}_1^{\left( 0 \right)} - \bnabla p_1^{\left( 0 \right)} = \IB{0},\\ [1 pt]
\; \;\, \quad\IB{v}_1^{\left( 0 \right)} = U_{sx}^{\left( 0 \right)}{\IB{e}_x} + \Omega _{sy}^{\left( 0 \right)}{\IB{e}_y} \times \IB{r} + H\!a{Z_p}(\bnabla {\psi _1} + \bnabla {\psi _2})\\
\; \, \quad \qquad \quad - \left( {\alpha  + \beta z + \gamma {z^2} + H\!a({Z_p} - {Z_w})} \right){\IB{e}_x} \qquad \quad \mbox{at\ } r = 1,\\
\; \; \quad\IB{v}_1^{\left( 0 \right)} \to \IB{0}\qquad \mbox{at\ }r \to \infty .
\end{array} \right\}
\label{Vel1}
\end{equation}
 \begin{equation}
  \left. \begin{array}{l}
 \bnabla  \bcdot \IB{v}_2^{\left( 0 \right)} = 0,\quad {\nabla ^2}\IB{v}_2^{\left( 0 \right)} - \bnabla p_2^{\left( 0 \right)} = \IB{0},\\ [1 pt]
\; \; \, \quad \IB{v}_2^{\left( 0 \right)} = H\!a{Z_w}\left( {\bnabla {\psi _1} + \bnabla {\psi _2}} \right) - \IB{v}_1^{\left( 0 \right)}\quad \mbox{at the walls,\ }\\
 \;  \; \, \quad \IB{v}_2^{\left( 0 \right)} \to \IB{0} \quad \mbox{at\ } r \to \infty .
 \end{array} \right\}
 \label{Vel2}
 \end{equation}
 \begin{equation}
 \left. \begin{array}{l}
 \bnabla  \bcdot \IB{v}_3^{\left( 0 \right)} = 0,{\nabla ^2}\IB{v}_3^{\left( 0 \right)} - \bnabla p_3^{\left( 0 \right)} = \IB{0},\\ [1 pt]
 \; \; \, \quad\IB{v}_3^{\left( 0 \right)} =  - \IB{v}_2^{\left( 0 \right)} \quad \mbox{at\ } r = 1,\\
 \; \; \, \quad \IB{v}_3^{\left( 0 \right)} \to \IB{0} \quad \mbox{at\ } r \to \infty .
 \end{array} \right\}
 \label{Vel3}
 \end{equation}\\
 
\textit{ Solution of }$ \IB{v}_{1}^{(0)} $: The odd reflections are found using Lamb’s general method \citep{lamb}. 
  We obtain the following leading terms for $ \IB{v}_{1}^{(0)} $:
\begin{equation}
\IB{v}_1^{\left( 0 \right)} = {A_1}\left( {{\IB{e}_x} + \frac{{x\IB{r}}}{{{r^2}}}} \right)\frac{1}{r} + {B_1}\left( { - {\IB{e}_x} + \frac{{3x\IB{r}}}{{{r^2}}}} \right)\frac{1}{{{r^3}}} + {C_1}\left( {\frac{{z{\IB{e}_x}}}{{{r^3}}} - \frac{{x{\IB{e}_z}}}{{{r^3}}}} \right) + {D_1}\left( {\frac{{zx\IB{r}}}{{{r^5}}}} \right).
\label{VelSol1}
\end{equation}
The coefficients $ A_{1},\:B_{1},\:C_{1} \:\mbox{and\ } D_{1} $ are associated with the stokeslet, source-dipole,rotlet (generated due to relative rotation of the particle) and stresslet flow disturbances, respectively. These are:
\begin{equation}
\begin{array}{l}
{A_1} = \frac{3}{4} ( U_{s x}^{\left( 0 \right)} - \alpha  - \frac{\gamma }{3} - H\!a({Z_p} - {Z_w}) + {\kappa ^3}H\!a{Z_p}\: \sl{2}),\\ [5 pt]
{B_1} = \frac{1}{2} (H\!a{Z_p}) - \frac{1}{4} ( U_{s x}^{\left( 0 \right)} - \alpha  - \frac{{3\gamma }}{5} - H\!a({Z_p} - {Z_w}) + {\kappa ^3}H\!a{Z_p}\: \sl{2}),\\ [5 pt]
{C_1} = \Omega _{s y}^{\left( 0 \right)} - \beta /2 ,\: \mbox{and\ }{D_1} =  - 5\beta /2.
\end{array}
\label{VelCoeff1}
\end{equation}
Here, $ \SL{2} $ is a vector denoting the contribution to slip at the particle surface due to wall correction of the disturbance potential. This correction enters the solution at $ O(\kappa^{3}) $.
 \begin{equation}
\SL{2} = \int\limits_0^\infty {{- \frac{1}{4}\left( {{b_2} + {b_3}} \right){\lambda ^2}} \mathrm{d}} \lambda \; \IB{e}_{x}.
\label{SlipCorr}
\end{equation}

\textit{Solution to }$ \IB{v}_{2}^{(0)} $: 
%The solution to wall reflected disturbance velocity is determined by the form of non-homogeneity in the boundary condition at the walls (see (\ref{3.21})). As in \S \ref{sec:3}.1, the non-homogeneities ($ H\!a{Z_w}{\bnabla {\psi _1},\: H\!a{Z_w} \bnabla {\psi _2}} \: \mbox{and\ } \IB{v}_1^{(0)} $) must be converted into the outer scale before applying Faxen’s integral transformation. It can be observed that  $ \psi_{2} $ is already defined in the integral form (see (\ref{3.11})), whereas $ \psi_{1}  $ and $ \IB{v}_{1}^{(0)} $ are defined in the particle scale ((\ref{3.4}) and (\ref{3.24}) respectively), yet to be transformed. 
%Upon performing Faxen transform \hl{we find that $ \bnabla \psi_{2} $ has a different integral form than $ \IB{v}_{1}^{(0)} $ and $ \bnabla \psi_{1} $}. 
%Therefore due to difference in the integral forms, we use superposition and seek $ \tilde{\IB{v}}_{2}^{(0)} $ as $ \tilde{\IB{v}}_{2\,(i)}^{(0)}+\tilde{\IB{v}}_{2\,(ii)}^{(0)} $. These components satisfy the following boundary conditions:
%
%\begin{equation}
%\tilde{\IB{v}}_{2\,(i)}^{(0)} = H\!a{{Z}_w}\kappa \tilde{\bnabla} (\tilde{\psi_{1}}) - \tilde{\IB{v}}_{1}^{(0)} \quad \mbox{at the walls,\ }
%\label{3.26}
%\end{equation}
%\begin{equation}
%\tilde{\IB{v}}_{2\,(ii)}^{(0)} = H\!a{{Z}_w}\kappa \tilde{\bnabla} (\tilde{\psi_{2}}) \quad \mbox{at the walls.\ }
%\label{3.27}\\ [5 pt]
%\end{equation}
Similar to  \S \ref{sec:3}.1, we use Faxen transformation technique to obtain the solution to $\tilde{\IB{v}}_{2}^{(0)}=\{\tilde{u}_{2}^{(0)},\:\tilde{v}_{2}^{(0)},\:\tilde{w}_{2}^{(0)}\} $ as:
\begin{equation}
\tilde u_{2}^{(0)} = \frac{1}{{2\pi }}\int_{ - \infty }^{ + \infty } {\int_{ - \infty }^{ + \infty } {{\mathrm{e}^{\mathrm{i}\Theta }}\left( \begin{array}{l}
		{\mathrm{e}^{(\frac{{ - \lambda Z}}{2})}}\left( {{\ell _4} + \frac{{{\xi ^2}}}{{{\lambda ^2}}}\left( {{\ell _5} + \frac{{\lambda Z}}{2}{\ell _6}} - {\frac{H\!a Z_{w} \kappa^{3} \lambda}{2} b_{2}} \right)} \right)\\
		{\rm{            }} + {\mathrm{e}^{(\frac{{ + \lambda Z}}{2})}}\left( {{\ell _7} + \frac{{{\xi ^2}}}{{{\lambda ^2}}}\left( {{\ell _8} - \frac{{\lambda Z}}{2}{\ell _9}} - {\frac{H\!a Z_{w} \kappa^{3} \lambda}{2} b_{3}} \right)} \right)
		\end{array} \right)\mathrm{d}\xi } } \mathrm{d}\eta ,
\label{VelSol2i}
\end{equation}
\begin{equation}
\tilde v_{2}^{(0)} = \frac{1}{{2\pi }}\int_{ - \infty }^{ + \infty } {\int_{ - \infty }^{ + \infty } {{\mathrm{e}^{\mathrm{i}\Theta }}\left( \begin{array}{l}
		{\mathrm{e}^{(\frac{{ - \lambda Z}}{2})}}\left( {{\ell _5} + \frac{{\lambda Z}}{2}{\ell _6}} - {\frac{H\!a Z_{w} \kappa^{3} \lambda}{2} b_{2}} \right)\\
		{\rm{                  }} + {\mathrm{e}^{(\frac{{ + \lambda Z}}{2})}}\left( {{\ell _8} - \frac{{\lambda Z}}{2}{\ell _9}} - {\frac{H\!a Z_{w} \kappa^{3} \lambda}{2} b_{3}} \right)
		\end{array} \right)\left( {\frac{{\xi \eta }}{{{\lambda ^2}}}} \right)\mathrm{d}\xi } } \mathrm{d}\eta,
\label{VelSol2ii}
\end{equation}
\begin{equation}
\tilde w_{2}^{(0)} = \frac{1}{{2\pi }}\int_{ - \infty }^{ + \infty } {\int_{ - \infty }^{ + \infty } {{\mathrm{e}^{\mathrm{i}\Theta }}\left( \begin{array}{l}
		{\mathrm{e}^{(\frac{{ - \lambda Z}}{2})}}\left( {{\ell _4} + {\ell _5} + \left( {1 + \frac{{\lambda Z}}{2}} \right){\ell _6}} - {\frac{H\!a Z_{w} \kappa^{3} \lambda Z}{2} b_{2}} \right)\\
		{\rm{      }} - {\mathrm{e}^{(\frac{{ + \lambda Z}}{2})}}\left( {{\ell _7} + {\ell _8} + \left( {1 - \frac{{\lambda Z}}{2}} \right){\ell _9}} - {\frac{H\!a Z_{w} \kappa^{3} \lambda Z}{2} b_{3}} \right)
		\end{array} \right)\left( {\frac{{\mathrm{i}\xi }}{\lambda }} \right)\mathrm{d}\xi } } \mathrm{d}\eta. \\ [5 pt]
\label{VelSol2iii}
\end{equation}
 Here, the terms $ {\ell _4},{\ell _5}\cdots,{\ell _9} $ are functions of the Fourier variable $ \lambda $ and the coefficients ($ A_{1},\;B_{1},\;C_{1}, \mbox{ and\ } D_{1} $) defined in (\ref{VelCoeff1}). Details of the derivation are skipped for brevity and can be found in the supplementary material.

At the leading order it suffices to represent the velocity disturbance field using the first two reflections. However, the solution is incomplete because the translational and angular velocities are unknown (see (\ref{VelCoeff1})). These can be found by imposing force-free and torque-free conditions on the particle at $ O(\Rey ^{0}) $: $ \IB{F}_{H}^{(0)} + \IB{F}_{M} = \IB{0} \mbox{ and\ } \IB{L}_{H}^{(0)} + \IB{L}_{M} = \IB{0} $. 
Since the electric force (\ref{FM}) acts only along the z-axis and the electric torque (\ref{LM}) is zero, $ U_{s\,x}^{(0)} $ and $ \Omega_{s\,y}^{(0)} $ are found by hydrodynamic force and torque balance in x and y directions, respectively.

\subsection{Evaluation of $ \IB{U}_{s}^{(0)} \mbox{and\ } \IB{\Omega}_{s}^{(0)} $}

For a particle-wall system, \citet[p. 239]{happel2012low} derived the hydrodynamic force and torque on a spherical particle in terms of successive reflections: 
\begin{equation}
\IB{F}_H^{\left( 0 \right)} = \IB{F}_{H\,1}^{\left( 0 \right)} + \IB{F}_{H\,3}^{\left( 0 \right)} +  \cdots
 \mbox{ and\ } \IB{L}_H^{\left( 0 \right)} = \IB{L}_{H\,1}^{\left( 0 \right)} + \IB{L}_{H\,3}^{\left( 0 \right)} +  \cdots.
\label{FHLH}
\end{equation}
Since the particle is absent in the formulation of even-numbered reflections, the hydrodynamic drag and torque, due to even fields vanishes. Thus, the contribution to force and torque arises only from the odd-numbered reflections.
%Owing to (\ref{3.21}), the force due to even fields vanishes: 
%\begin{equation}
%\IB{F}_{H\,2m}^{\left( 0 \right)} = \int\limits_{{S_p}} {\IB{\sigma} _{H\,2m}^{(0)} \bcdot \IB{n}\,\mathrm{d}S}  = \int\limits_{{V_f}} {\left( {\bnabla  \bcdot \IB{\sigma} _{H\,2m}^{(0)}} \right)\,\mathrm{d}V}  = \IB{0} \mbox{  , m is a positive integer.}
%\label{3.35}
%\end{equation}
 
 At the current order of reflections of the velocity disturbance field ($ \IB{v}^{(0)}=\IB{v}^{(0)}_{1} + \IB{v}^{(0)}_{2} $), imposing force-free and torque-free conditions provides translational and rotational velocities without the wall correction. $ \IB{F}_{H\,3}^{(0)} $ incorporates the first order wall effects into $ U_{s\,x}^{(0)} $ and $ \Omega_{s\,y}^{(0)} $, which requires $ \IB{v}_{3}^{(0)} $. Appendix \ref{appA}.4 describes the determination of $ \IB{v}_{3}^{(0)} $ using Lamb's general solution \citep{lamb}.
 
The force and torque on a spherical particle can be expressed through the coefficients of stokeslet and rotlet disturbances, respectively. Following \citet[p. 88]{kim2013}, we write (\ref{FHLH}) as:
\begin{equation}
F_{H\,x}^{\left( 0 \right)} =  - 4\pi \left( {{A_1} + {A_3} +  \cdots } \right)
\: \mbox{and\ } L_{H y}^{\left( 0 \right)} =  - 8\pi \left( {{C_1} + {C_3} +  \cdots } \right).
\label{FHLHAC}
\end{equation}
The coefficients $ A_{1},\: C_{1} $ and $ A_{3},\: C_{3} $ are associated with the Lamb’s solution of the first reflection of the velocity disturbance (\ref{VelSol1}) and third reflection (see \ref{A5}), respectively. $ A_{3} \mbox{ and\ } C_{3} $ can be expressed in terms of  $ A_{1},\:B_{1},\:C_{1} \mbox{ and\ } D_{1} $ as:
\begin{equation}
A_{3}={A_1}\left( {\kappa {W_A}} \right) + {B_1}\left( {{\kappa ^3}{W_B}} \right) + {C_1}\left( {{\kappa ^2}{W_C}} \right) + {D_1}\left( {{\kappa ^2}{W_D}} \right)
  - \frac{3}{4}{\kappa ^3}Ha{Z_w}\: \sl{2},
\label{A3}
\end{equation}
\begin{equation}
 {C_3} = {A_1}\left( {{\kappa ^2}{{\cal X}_A}} \right) + {B_1}\left( {{\kappa ^3}{{\cal X}_B}} \right) + {C_1}\left( {{\kappa ^4}{{\cal X}_C}} \right).\\ [5 pt]
\label{C3}
\end{equation}
Here, $ \kappa {W_A},\:{\kappa ^3}{W_B},\:{\kappa ^2}{W_C}\:\mbox{and\ }{\kappa ^2}{W_D} $ represent the wall corrections to hydrodynamic drag due to the reflection of stokeslet, source-dipole, rotlet and stresslet disturbances, respectively. Similarly, $ \kappa^{2} {\cal{X}_A},\:{\kappa ^3}{\cal{X}_B},\:\mbox{and\ }{\kappa ^4}{\cal{X}_C} $ represent wall correction to the hydrodynamic torque. These corrections are integrals over Fourier variable $ \lambda $ and are a function of distance from the walls.\\

Substitution of (\ref{A3}) and (\ref{C3}) into (\ref{FHLHAC}) results in a system of two equations for: $ U_{s\,x}^{(0)} $ and $ \Omega_{s\,y}^{(0)} $. Imposing the hydrodynamic force and torque to be zero and upon expanding the coefficients $ A_{1},\:B_{1},\:C_{1},\: \mbox{and\ } D_{1} $, we obtain:
\begin{equation}
  U_{s\,x}^{(0)} \approx \left\{ \begin{array}{l}
\left[\,( {\alpha  + {\gamma  \mathord{\left/{\vphantom {\gamma  3}} \right.\kern-\nulldelimiterspace} 3}} \right)\left( {1 + \kappa W_A} \right) - {{10{\kappa ^2}\beta W_D} \mathord{\left/{\vphantom {{10{\kappa ^2}\beta W_D} 9}} \right.\kern-\nulldelimiterspace} 9}\,]/(1 + \kappa W_A)\\ [5 pt]

\; + H\!a\left( {{Z_p} - {Z_w}} \right)\left( {1 - \frac{2}{3}\frac{{W_B{\kappa ^3}}}{{\left( {1 + \kappa W_A} \right)}}} \right) - {\kappa ^3}\: \sl{2} H\!a\left( {{Z_p} - \frac{{{Z_w}}}{{\left( {1 + \kappa W_A} \right)}}} \right),
\end{array} \right.
\label{U0}
\end{equation}
\begin{equation}
\Omega_{s\,y}^{(0)} \approx \frac{{\beta \left( {1 + \kappa W_A} \right) - {{10{\cal X}_D{\kappa ^3}\beta } \mathord{\left/
				{\vphantom {{10{\cal X}_D{\kappa ^3}\beta } 3}} \right.
				\kern-\nulldelimiterspace} 3}}}{{2\left( {1 + \kappa W_A} \right)}} - H\!a\left( {{Z_p} - {Z_w}} \right)\frac{{{\cal X}_B{\kappa ^4}}}{{2\left( {1 + {\cal X}_C{\kappa ^3}} \right)}}.
\label{Om0}
\end{equation}\vspace{3mm}
Validation of the above two velocities with the literature can be found in Appendix \ref{appB}.

With (\ref{VelSol1}), (\ref{VelSol2i})-(\ref{VelSol2iii}), (\ref{U0}) and (\ref{Om0}), we have obtained the solution to $ \IB{v}^{(0)} $. We use a similar procedure to evaluate the test field $ \IB{u}^{t} $ governed by (\ref{Test}) (the details are described in Appendix \ref{appA}.5). The calculation of the inertial lift force using the disturbance and test field is discussed next.

 \section{Inertial force volume integral}\label{sec:4}
 
%From (\ref{3.17}) we found that the electrical force ($ \IB{F}_{M} $) acts in the lateral direction. . 
In \S \ref{sec:3}.1, we found that the electrical force $ \IB{F}_{M} $ acts in the lateral direction (see (\ref{FM})).
The total lateral force experienced by an electrophoretic particle arises from inertial stresses ($ F_{I} $) and Maxwell stresses ($ F_{M\,z} $).
Since the primary objective of this work is to determine lateral migration due to inertia, we now focus on $ F_{I} $. Later,\ in section \S\ref{sec:9}, we compare the contributions from $ F_{I} $ and $ F_{M \, z} $. 
	
 In \S \ref{sec:2}, we derived the inertial migration force (\ref{MigF}) as:
 \begin{equation}
 F_{I}=-\Rey_{p} \int_{{V_f}} \IB{u}^{t} \bcdot \IB{f}\; \mathrm{d}V.
 \label{4.1}
 \end{equation}
  Here, $ \IB{f}={{\IB{v}^{(0)}} \bcdot \bnabla {\IB{v}^{(0)}} + \IB{V}_\infty ^{(0)} \bcdot \bnabla {\IB{v}^{(0)}} + {\IB{v}^{(0)}} \bcdot \bnabla \IB{V}_\infty ^{(0)}} $. Before discussing the evaluation of (\ref{4.1}), in this section, we draw a few qualitative insights on the behaviour of the lift force integral.
  
 The domain of integration extends from the surface of particle to the walls of the channel. % Since the particle is spherical in shape and walls are rectangular (infinite in x and y directions), it is convenient to divide the fluid domain into two sub-domains 
Following
  \citet{ho1974,hood2015}, we divide the fluid domain: $ V_{f}=V_{1}+V_{2} $. These sub-domains are defined by:
 \begin{equation}
 {V_1} = \left\{ {\IB{r} : 1 \le |r| \le \delta } \right\},\quad {V_2} = \left\{ {\IB{R} : \kappa \delta  \le |R| \le \infty } \right\}.
 \label{4.2}
 \end{equation}
 Here $ \delta $ is an intermediate radius which satisfies: $ 1 \ll \delta\ll 1/\kappa $. The length scale corresponding to the inner domain ($ V_{1} $) is particle radius ‘\textit{a}’. Whereas, channel width ‘\textit{l}’ is the characteristic length in the outer domain ($ V_{2} $).
 
 We first inspect and compare the order of magnitude of contributions arising from the volume integrals in the inner and outer domains. \citet{ho1974} and \citet{hood2015}, in the absence of electrophoresis, found that the contribution from inner integral was O($ \kappa $) smaller than that from the outer integral. \citet{ho1974} hence neglected the contribution from the inner integral. In this section, we show that such a simplification is not justified for the electrophoretic case.

 \subsection{Inner integral}
 The order of magnitude of the velocity fields in terms of size ratio $ \kappa $ and radial distance $ r $, at the leading order are:
 \begin{equation}
 \IB{v}_{1}^{(0)} \sim O(1/r^{3})H\!a\,Z_{p} + O(\kappa/r^{2}),\quad \IB{v}_{2}^{(0)} \sim O(\kappa^{3})+O(\kappa^{3})H\!a\,(Z_{p}-Z_{w}),
 \label{4.3}
 \end{equation}
 \begin{equation}
 \IB{u}_{1}^{(t)} \sim O(1/r)+O(1/r^{3}),\quad \IB{u}_{2}^{(t)} \sim O(\kappa)+O(\kappa^{3}). \\ [5 pt]
 \label{4.4}
 \end{equation}
 Substituting the translational velocity (\ref{U0}) into the undisturbed background flow ($ \IB{V}_{\infty}^{(0)} = \left( {\alpha  + \beta z + \gamma {z^2}} \right){\IB{e}_x} + H\!a{{Z}_w}\IB{\nabla} {\phi _\infty } - {\IB{U}_s^{(0)}} $), we obtain the order of magnitude of $ \IB{V}_{\infty}^{(0)} $ as:
 \begin{equation}
 \IB{V}_\infty ^{(0)} \sim O\left( {\kappa r} \right) + O\left( {{\kappa ^2}{r^2}} \right) + O\left( 1 \right)H\!a{Z_p} +  \cdots.
 \label{4.5}
 \end{equation}
Using (\ref{4.3})-(\ref{4.5}) in (\ref{4.1}), we obtain the integrand as:
 \begin{eqnarray}
% \begin{array}{l}
 \IB{u}^{t} \bcdot \IB{f} & \sim & O\left( {\frac{{{{\left( {H\!a{Z_p}} \right)}^2}}}{{{r^5}}} + \frac{{{{\left( {H\!a{Z_p}} \right)}^2}}}{{{r^7}}}+ \cdots} \right) + \kappa \, O\left( {\frac{{H\!a{Z_p}}}{{{r^4}}} + \frac{{H\!a{Z_p} }}{{{r^6}}} +  \cdots } \right) \nonumber  \\ 
 && + {\kappa ^2}O\left( {\frac{1}{{{r^3}}} + \frac{{H\!a{Z_p}}}{{{r^5}}}  +  \cdots } \right) + {\kappa ^3}O\left( {\frac{1}{{{r^2}}} +\frac{1}{{{r^4}}} +  \cdots } \right) +  \cdots .
 \label{4.6}
 \end{eqnarray}
 The volume integral of the first term is identically zero. The leading order contribution arises from the second $ O(\kappa) $ term:
\begin{equation}
\mathop {\lim }\limits_{\delta  \to \infty } {F_{V1}} =  - \Rey{_p}\left( {0 + \left( {\frac{{47\pi }}{{20}}H\!a{{Z}_p}} \right)\beta  + 0 + O\left( {{\kappa ^3}} \right) +  \cdots } \right).
\label{4.7}
\end{equation}
In the above expression, the proportionality of $ H\!a\,Z_{p} $ suggests that the $ O(\kappa) $ contribution (implicit in background shear $ \beta $) arises due to electrophoresis. The contribution due to stresslet disturbances and wall reflections is $ O(\kappa^{3}) $. For electric field in the same direction as the flow ($ H\!a >0 $), a positive $ Z_{p} $ corresponds to a leading particle. Below centerline, the shear ($ \beta $) is positive and the overall expression in (\ref{4.7}) is therefore negative. This corresponds to a lift in negative z-direction which pushes the leading particle towards the wall. A similar explanation can be used to show that a lagging particle ($ Z_{p}<0 $) experiences a lift (\ref{4.7}) towards the center. These effects were also experimentally observed by \citet{kim2009,kim2009Three,cevheri2014,yuan2016,xuan2018,yee2018}. Expression (\ref{4.7}) also reveals that the wall zeta potential does not affect the lift because its contribution to the disturbance is $ O(\kappa^{3}) $ (see (\ref{4.3})). 

In the absence of electrophoresis, the inner integral lift contributes at $ O(\kappa^{3}) $, as also previously reported by \citet{ho1974} and \citet{hood2015}.
Next we estimate the contribution arising from the outer integral.

\subsection{Outer integral}
Since the walls lie at distances $ O(1/\kappa) $, the velocity field expressions are rescaled with respect to R$ (=\kappa r) $ (similar to (\ref{Rescale})). 
The inertial lift force expression in the outer coordinates is: 

$ \begin{aligned}
{F_{V2}} =  - \Rey{_p}\int\limits_{{V_2}} {\left( {\tilde{\IB{u}}_1^t + \tilde{\IB{u}}_2^t} \right) \bcdot \left[ \begin{array}{l}
	\left( {\tilde{\IB{v}}_1^{\left( 0 \right)} + \tilde{\IB{v}}_2^{\left( 0 \right)}} \right) \bcdot \kappa \tilde \bnabla \left( {\tilde{\IB{v}}_1^{\left( 0 \right)} + \tilde{\IB{v}}_2^{\left( 0 \right)}} \right) + \\
	\tilde{\IB{V}}_\infty ^{(0)} \bcdot \kappa \tilde \bnabla \left( {\tilde{\IB{v}}_1^{\left( 0 \right)} + \tilde{\IB{v}}_2^{\left( 0 \right)}} \right) + \left( {\tilde{\IB{v}}_1^{\left( 0 \right)} + \tilde{\IB{v}}_2^{\left( 0 \right)}} \right) \bcdot \kappa \tilde \bnabla \tilde{\IB{V}}_\infty ^{(0)}
	\end{array} \right]}{\kappa ^{ - 3}} \mathrm{d}\tilde V
\end{aligned} $
\begin{equation}
\begin{aligned}
= - {{\mathop{\rm Re}\nolimits} _p}\,{\kappa ^{ - 2}}\int\limits_{{V_2}} {{{\tilde{\IB{u}}}^t} \bcdot \tilde{\IB{f}}} \,  \mathrm{d}\tilde V.
\end{aligned}
\label{4.8}
\end{equation}
In the outer coordinate representation, the order of magnitude of the velocity fields in terms of size ratio $ \kappa $, at the leading order are:
\begin{equation}
\tilde{\IB{v}}_{1}^{(0)} \sim O(\kappa^{3})+O(\kappa^{3})Ha\,Z_{p},\quad \tilde{\IB{v}}_{2}^{(0)} \sim O(\kappa^{3})+O(\kappa^{3})Ha\,(Z_{p}-Z_{w}),
\label{4.9}
\end{equation}
\begin{equation}
\tilde{\IB{u}}_{1}^{(t)} \sim O(\kappa)+O(\kappa^{3}),\quad \tilde{\IB{u}}_{2}^{(t)} \sim O(\kappa)+O(\kappa^{3}).
\label{4.10}
\end{equation}
The order of magnitude of the undisturbed background velocity is:
\begin{equation}
{\tilde{\IB{V}}_\infty } \sim O\left( 1 \right) + O\left( 1 \right)Ha{Z_p} + O\left( {{\kappa ^2}} \right) +  \cdots.
\label{4.11}
\end{equation}
Substituting (\ref{4.9}), (\ref{4.10}) and (\ref{4.11}) in (\ref{4.8}), we find that the integrand is $ O(\kappa^{2}) $. 
Since the contribution from the inner and outer integrals are comparable, it is important that the integration must be carried out from the particle surface to the walls.

\subsection{Evaluation of volume integral}
The region of integration extends from the surface of the spherical particle  to the rectangular walls, which extend to infinity in the $ x $ and $ y $ directions. A direct numerical approach would be computationally intensive as the evaluation of the lift force involves numerous nested Fourier integrals, apart from integrals in $ x $, $ y $ and $ z $ coordinates
\footnote{Faxen transformation provides the wall reflected fields in terms of Fourier integrals, which makes it cumbersome to evaluate the lift force volume integral (as it consists 7 nested integrals). Using certain transformations we reduce this number to 5 (see supplementary information).}.
For computational convenience, we resort to a splitting of the region over which the integral is evaluated. 
We perform the integration in two sub-regions, where the region of integration extends from: (i) the spherical surface to a circumscribing cylinder and (ii) the cylindrical surface to the walls ($ z $-direction) and infinity ($ x $ and $ y $ directions). 
This simplifies the calculation as; in sub-region-(i) we employ a spherical coordinate system and evaluate the integral with the help of trigonometric identities; in sub-region-(ii) a cylindrical coordinate system is employed to perform the integration.
We evaluate the integral in both the sub-regions and include it in our results, which ensures no loss of accuracy.

We choose to evaluate the volume integral in the outer coordinates, over the entire volume ($ V_{f} $). For sub-region-(ii), the region of integration can be visualized as a cylinder $ ({Z_C} \in [ - s,1 - s] \mbox{ and\ } {R_C} \in [0,\infty )) $ excluding an asymptotically small cylindrical ‘cavity’ (radius $ \in [0,\kappa ]$, height $ \in [ - \kappa ,\kappa ]$). 
The velocities are transformed into cylindrical coordinates $ (R_{C},\:\theta_{C},\:Z_{C}) $. The lift force volume integral corresponding to sub-region-(ii) is:
\begin{equation}
{F_I} =  - {{\mathop{\rm Re}\nolimits} _p}\int\limits_0^{2\pi } {{\kappa ^{ - 2}}\left\{ \begin{array}{l}
	\int\limits_0^\infty  {\int\limits_{ - s}^{ - \kappa } {\tilde{\IB{u}}^{t} \bcdot \tilde{\IB{f}}\: {R_C}\:\mathrm{d}{Z_C}} \mathrm{d}{R_C}} \\ [3 pt]
	+ \int\limits_\kappa ^\infty  {\int\limits_{ - \kappa }^{ + \kappa } { \tilde{\IB{u}}^{t} \bcdot \tilde{\IB{f}} \: {R_C}\:\mathrm{d}{Z_C}} \mathrm{d}{R_C}} \\ [3 pt]
	{\rm{  }} + \int\limits_0^\infty  {\int\limits_\kappa ^{1 - s} {\tilde{\IB{u}}^{t}  \bcdot \tilde{\IB{f}}\: {R_C}\:\mathrm{d}{Z_C}} \mathrm{d}{R_C}} 
	\end{array} \right\}} \mathrm{d}{\theta _C}
\label{4.12}
\end{equation}
The integration in the above equation is carried out analytically in $ {\theta _C},\:{R_C}\:\mbox{and\ }{Z_C} $ directions. The integration over the Fourier parameters is performed numerically using Gauss-Kronrod quadrature inbuilt in Mathematica $ 11.0 $. %\footnote{A technique to overcome an apparent divergence, in evaluating the volume integral, is described in supplementary information S2}. 
The contribution from sub-region-(i) is evaluated analytically (see Appendix \ref{appD}) and incorporated in our results.\\

%%%%%%%%%%%%%%%%%%%%%%%%%%%%%%%%%%%%%%%%%%%%%%%%%%%%%%%%%%%%%%%%%%%%%%%%%%%%%%%%%%%%%%%%

\section{Summary of the procedure}\label{sec:5}
Here we summarize the solution procedure followed in this work for the evaluation of inertial lift volume integral (\ref{MigF}).

\begin{enumerate}[leftmargin=0.7cm]
	\item {\,}First, the electrostatic potential ($ \psi $) is evaluated using the method of reflections ($ \psi_{1}+\psi_{2} $) in \S\ref{sec:3}.1. $ \psi_{1} $ is found in terms of spherical harmonics, whereas the second reflection is obtained via Faxen transformations \citep{faxen1922}.
	
	\item {\,}The velocity disturbance field is obtained using the method of reflections in \S\ref{sec:3}.2. The first reflection ($ \IB{v}_{1}^{(0)} $) is found using Lamb's general solution \citep{lamb}. The second reflection is found using Faxen transformations. The third reflection is found by again making use of Lamb's solution.
	
	\item {\,}The unknown translational ($ \IB{U}_{s}^{(0)} $) and rotational ($ \IB{\Omega}_{s}^{(0)} $) velocities are found by imposing force-free and torque-free conditions on the particle in \S\ref{sec:3}.3.
	
	\item {\,} The test field is evaluated in Appendix \ref{appA}5, following the methodology in (ii).
	
	\item {\,}The evaluated fields are substituted into the inertial lift force integral (\ref{MigF}).

\end{enumerate}
In the following sections, we discuss the results for the inertial lift ($ F_{I} $) experienced by the particle.

%%%%%%%%%%%%%%%%%%%%%%%%%%%%%%%%%%%%%%%%%%%%%%%%%%%%%%%%%%%%%%%%%%%%%%%%%%%%%%%%%%%%%%%%

\section{Inertial migration of an inert particle}\label{sec:6}
The inertial migration of a neutrally buoyant inert particle is discussed in this section. We consider pure hydrodynamic flows (such as: Couette and Poiseuille flow) and understand the mechanisms which contribute to the lift by using an order-of-magnitude analysis.

\subsection{Validation with earlier theoretical studies}
\begin{figure}%
	\centering
	\sidesubfloat[]{{\includegraphics[scale=0.35]{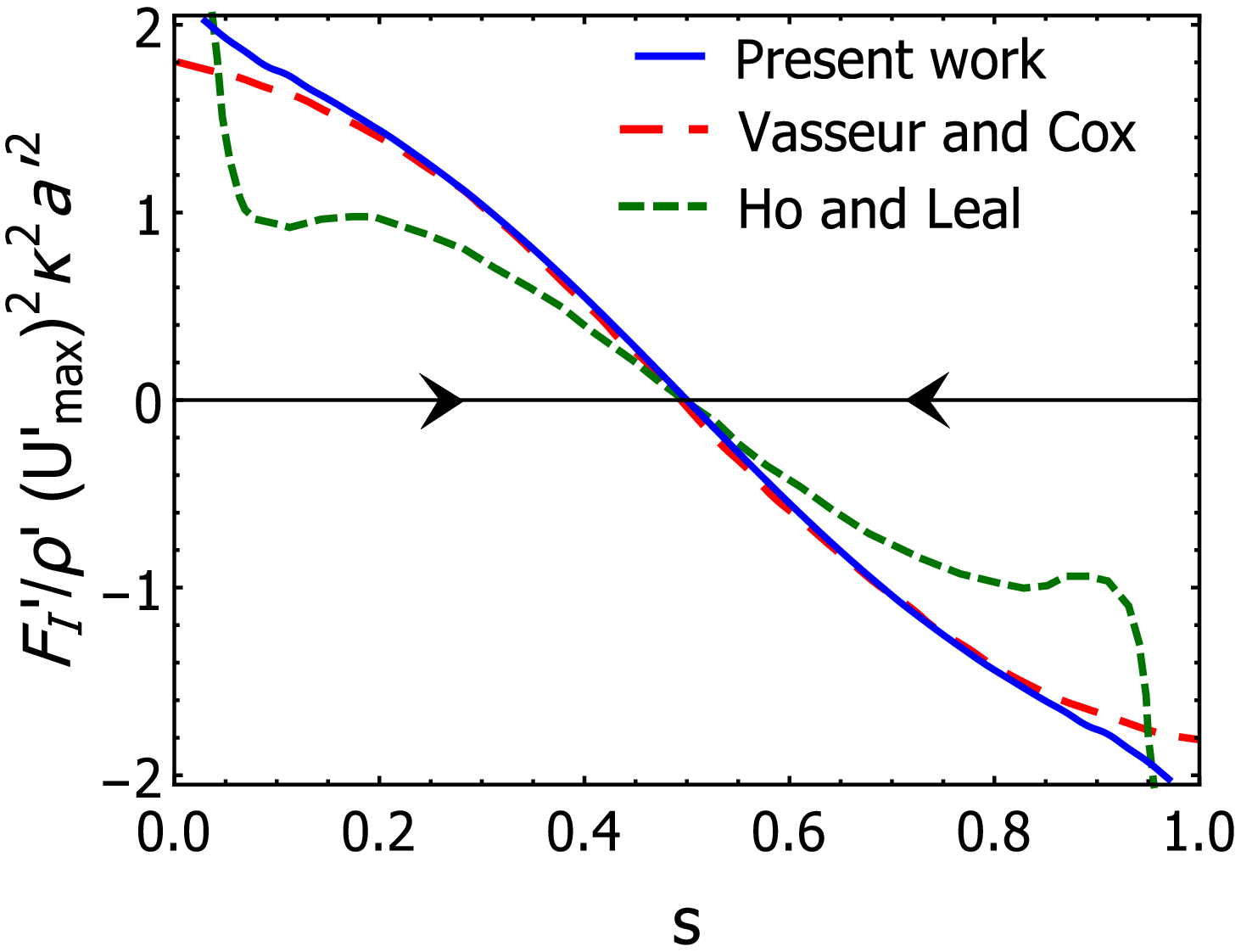} }}%
	\qquad
	\sidesubfloat[]{{\includegraphics[scale=0.35]{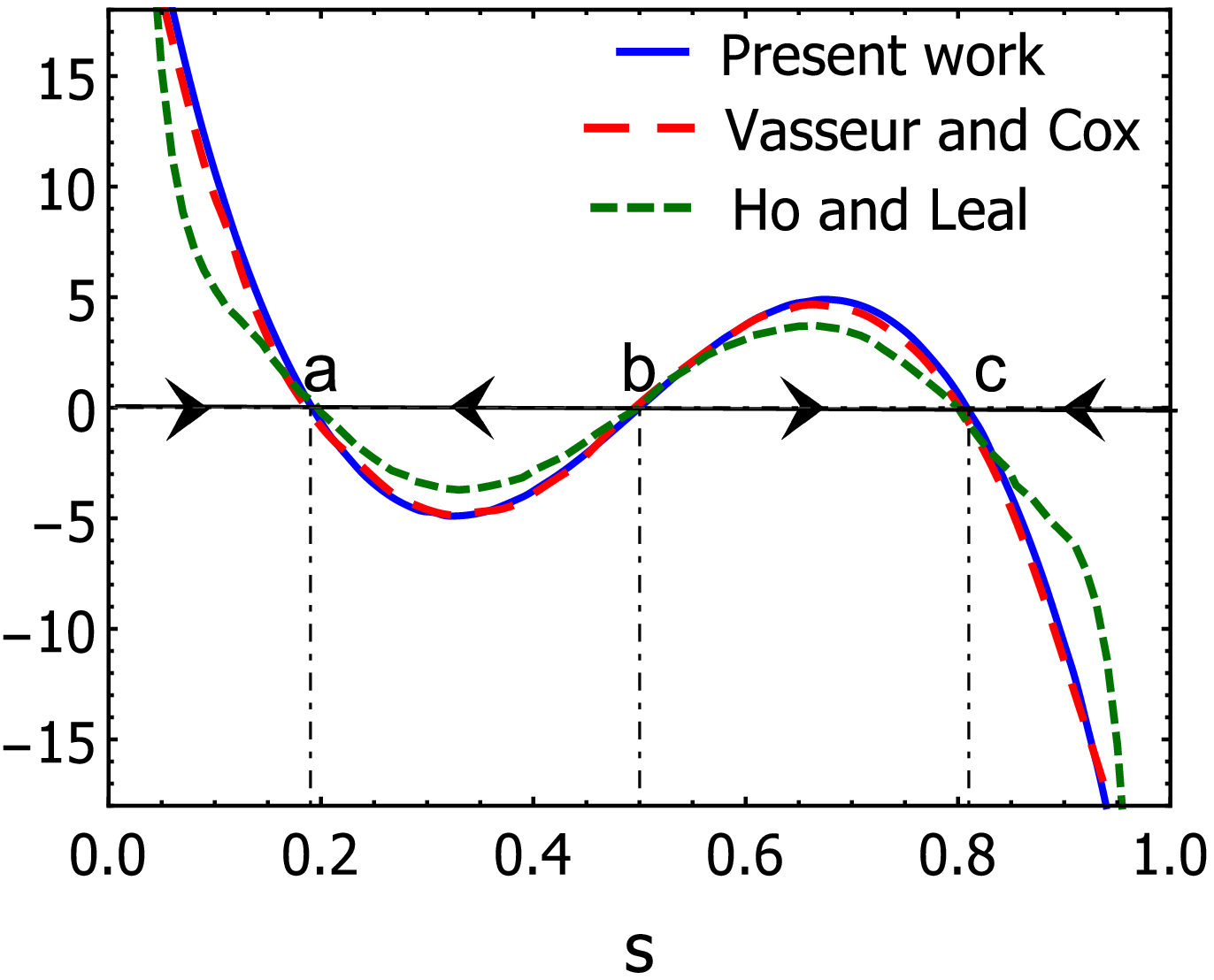} }}%
	\caption{Comparison with the previous theoretical studies \citep{ho1974,vasseur1976} for the inertial lift force experienced by a particle, as a function of distance from the bottom wall for: (a) Couette flow and (b) Poiseuille flow.}%
	\label{fig:4}%
\end{figure}

We first compare the results obtained in our work with those in the literature \citep{ho1974,vasseur1976}. Figure \ref{fig:4}(a) shows the lift force experienced by a neutrally buoyant sphere freely suspended in Couette flow. There exists a single equilibrium position at the center where the particle experiences no lift. Below the centerline ($ s<0.5 $) the force acting on the particle is positive, resulting in a lift in the positive z-direction (i.e. towards the centerline). Above the centerline ($ s>0.5 $) the lift force is in the negative z-direction (i.e. again towards the centerline). Hence, a neutrally buoyant particle in a Couette flow always migrates to the center of the channel (a stable equilibrium) as depicted by the arrows on the axis. 

For Poiseuille flow (figure \ref{fig:4}(b)), three equilibrium positions ‘a’, ‘b’ and ‘c’ exist ($ s \approx 0.19,\: 0.5,\: \mbox{and\ }0.81 $). Points ‘a’ and ‘c’ are characterized by a negative slope of the lift force curve and are therefore stable equilibrium positions, while point ‘b’ is unstable (arrows on the axis depict the direction of migration).

 Our results (for both Couette and Poiseuille flow) obtained by using Ho and Leal’s formulation are in good agreement with \citet{vasseur1976} who used a Green’s function technique to approximate the particle as a point force. Near the walls, lift force deviates significantly  from \citet{ho1974}. This may have been due to a failure in their numerical convergence near the walls, as pointed out by \citet{vasseur1976}.

\floatsetup[figure]{style=plain,subcapbesideposition=top}

\subsection{Insights on the origins of the inertial lift}
The approach adopted in this work allows us to find different contributions to the lift force on a particle suspended in a Poiseuille flow. These constituents are interpreted using an order-of-magnitude analysis. 

In the outer coordinates, the order of magnitude of background velocity is :
\begin{equation}
\tilde{\IB{V}}_\infty ^{\left( 0 \right)}\sim O\left( {{1 \mathord{\left/
			{\vphantom {1 \kappa }} \right.
			\kern-\nulldelimiterspace} \kappa }} \right)\beta  + O\left( {{1 \mathord{\left/
			{\vphantom {1 {{\kappa ^2}}}} \right.
			\kern-\nulldelimiterspace} {{\kappa ^2}}}} \right)\gamma  + O\left( 1 \right)\gamma ,
\label{6.1}
\end{equation}
the disturbance velocity (stresslet) is:
\begin{equation}
{\tilde{\IB{v}}^{\left( 0 \right)}}\sim O\left( {{\kappa ^2}} \right)\beta \left( {1 + {W_D}} \right),
\label{6.2}
\end{equation}
and the test velocity field is:
\begin{equation}
{\tilde{\IB{u}}^t}\sim O\left( \kappa  \right)\left( {1 + W_A^t} \right).
\label{6.3}
\end{equation}
 Here, $ W_{D} $ and $ W_{A}^{t} $ are the wall correction factors for stresslet and stokeslet (test field) disturbances, respectively. 
 
 Upon substitution of (\ref{6.1})-(\ref{6.3}) into the lift force expression (\ref{4.12}), we obtain:
 \begin{equation}
O\left( \frac{F_{I}}{\Rey_{p}} \right) \sim \int\limits_{{V_f}} {\left[ \, O\left( 1 \right)\beta \beta \left( {1 + {W_D}} \right) + O\left( {{1 \mathord{\left/
				{\vphantom {1 \kappa }} \right.
				\kern-\nulldelimiterspace} \kappa }} \right)\gamma \beta \left( {1 + {W_D}} \right) \right]\mathrm{d}\tilde V} .
\label{6.4}
\end{equation}
The first term in the above equation is the stresslet-wall force (known in the literature as the ‘wall lift’ force). It originates from the interaction of the stresslet disturbance with the wall. This results in a pressure asymmetry around the particle. The pressure is larger on the side facing the closer wall \citep{feng1994}. This force always acts away from the wall, even if the shear direction (the sign of $ \beta $) is reversed, as it is proportional to $ \beta^{2} $. The second term is the stresslet-curvature force (known in the literature as the ‘shear-gradient lift force’), arising from the interaction of the stresslet disturbance ($ \beta (1+W_{D}) $) with curvature in the background flow $ (\gamma) $. Owing to the proportionality to $ \beta \gamma $, this force always acts towards the walls.

\begin{figure}
	\centerline{\includegraphics[scale=0.4]{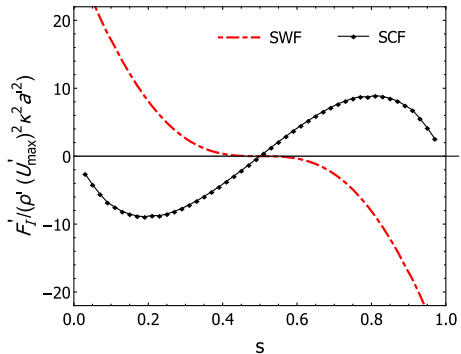}}% Images in 100% size
	\caption{Variation of leading constituents of inertial lift force as a function of distance from the bottom wall for Poiseuille flow. Here, stresslet-wall force (SWF) is the ‘wall lift force’ and stresslet-curvature force (SCF) is the ‘shear-gradient lift force’.}
	\label{fig:5}
\end{figure}

Figure \ref{fig:5} shows the variation of the above two components with the lateral position. 
The stresslet-wall force (SWF) acts to push the particle towards the centerline. Since SWF depends on the background shear ($ \beta^{2} $), it vanishes at the center and increases near the walls.
On the other hand, the stresslet-curvature force (SCF) has a destabilizing effect at the centerline i.e. a slight perturbation to the particle would push it away from the center. SCF increases (in magnitude) away from the centerline, and decreases near the stationary walls due to increased viscous resistance. 

From (\ref{6.4}) it can be observed that the direction of SCF is proportional to $ sgn(\beta \gamma) $.  If either the shear or the curvature direction is changed, the direction of SCF can reverse. This can be exploited to control the positions at which the particle gets focused. \citet{lee2018} recently showed that the curvature of the background flow profile can be controlled by using stratified flows with different viscosities. By varying the viscosity ratio, an external control over the focusing position was experimentally demonstrated.

%%%%%%%%%%%%%%%%%%%%%%%%%%%%%%%%%%%%%%%%%%%%%%%%%%%%%%%%%%%%%%%%%%%%%%%%%%%%%%%%%%%%%%%%

\section{Electrophoresis combined inertial lift}\label{sec:7}

\begin{figure}%
	\centering
	\sidesubfloat[]{{\includegraphics[scale=0.287]{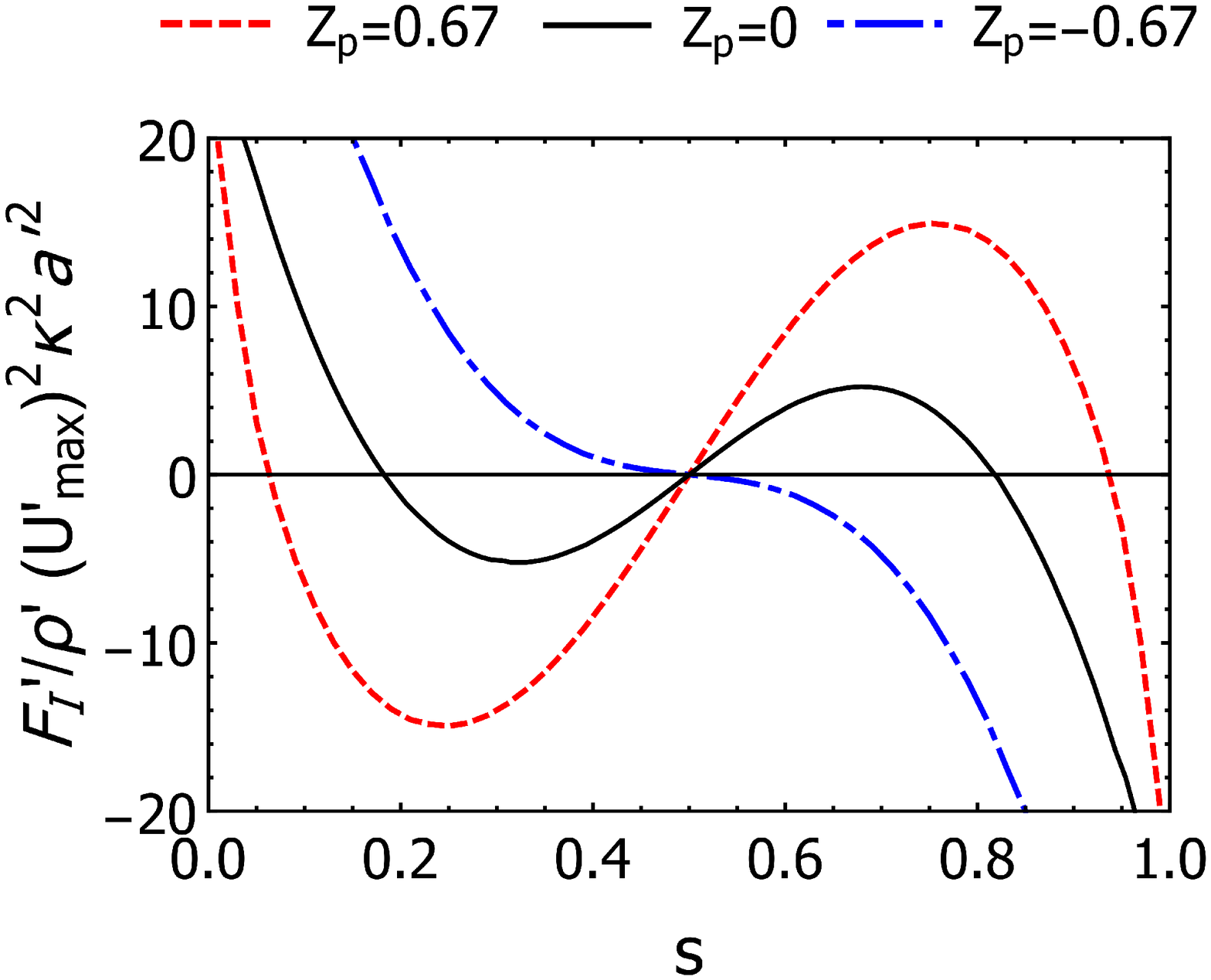} }}%
	\qquad
	\sidesubfloat[]{{\includegraphics[scale=0.297]{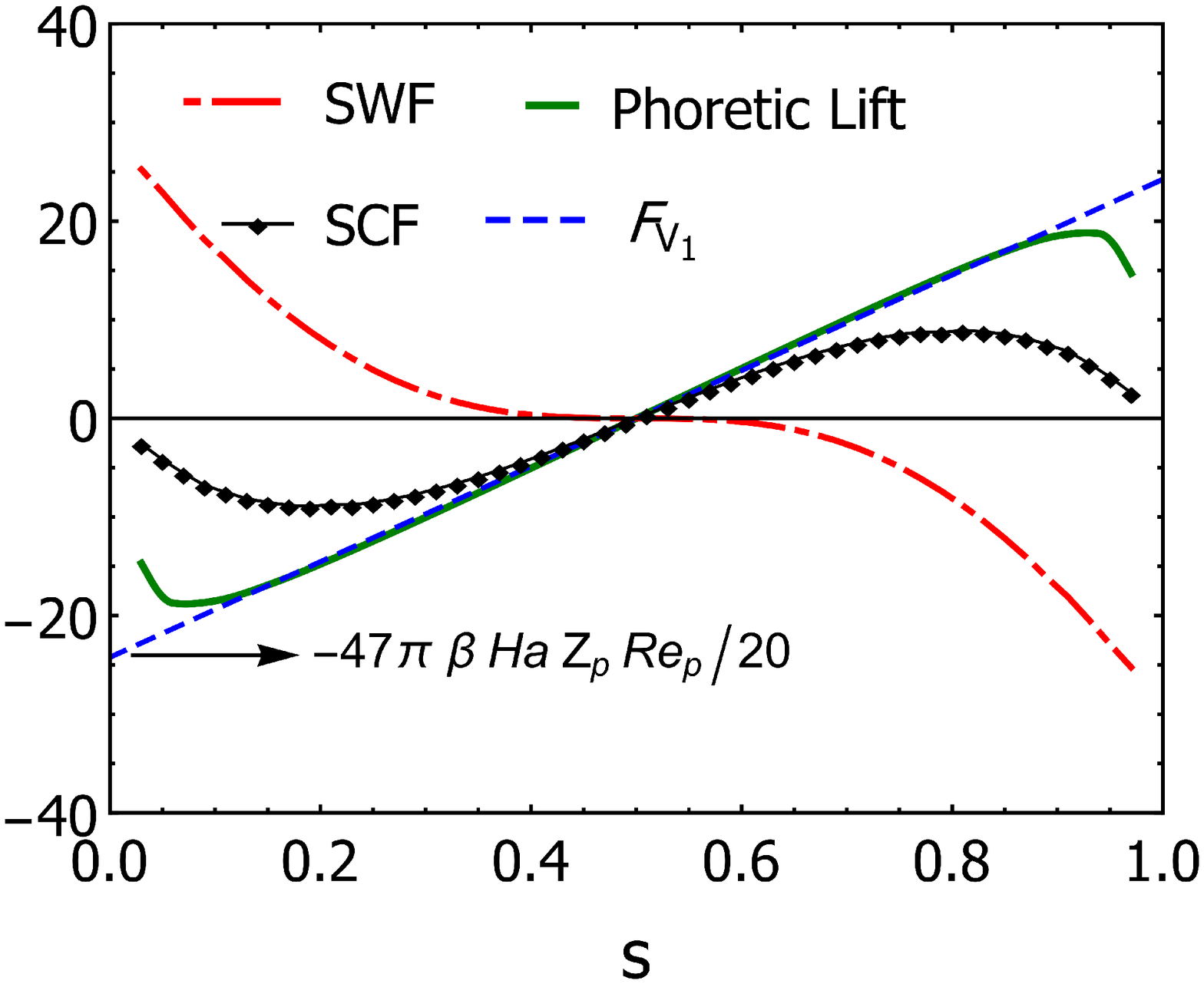} }}%
	\caption{Variation of the inertial lift force on the electrophoretic particle with distance from the bottom wall: (a) Effect of Zeta potential for $ H\!a= $1.25 ($ E_{\infty}'= $70 V$ / $cm); (b) Comparison of the phoretic-lift with SCF (stresslet-curvature force) and SWF (stresslet-wall force) for $ Z_{p}=0.67 $. The dashed line shows the analytical expression, expressed in (\ref{4.7}) (which does not account for the wall effects). Parameters: $ \zeta_{p}'=47\: \mbox{mV,\ } U_{max}'=2.25\: \mbox{mm/s,\ } a'=10\: \mu \mbox{m,\ } l'=1 \mbox{\:mm,\ } \varepsilon'=80\,\varepsilon_{0}' \mbox{,\ } \mu'=10^{-3} \mbox{\:Pa\,s.\ }  $ }%
	\label{fig:6}%
\end{figure}

We now focus on the effect that electrophoresis has on the inertial lift. This effect is captured in the absence of walls by the analytical expression derived in \S \ref{sec:4}.1 (see (\ref{4.7})). Here, we discuss the results which incorporate the wall effects to first order. In figure \ref{fig:6}(a), the variation of the lift force with position is shown for three different systems in which the electric field is applied in the same direction as the flow. A particle with $ Z_{p}=0 $ corresponds to an inert particle and has three equilibrium positions as discussed in \S \ref{sec:6}.1. A particle with $ (Z_{p}>0) $, leads the background flow, and the stable equilibrium position shifts towards the walls. For a particle with $ Z_{p}<0 $, the stable equilibrium positions shift inwards towards the center. This shift in equilibrium positions based on zeta potential is predicted by equation (\ref{4.7}). These trends are consistent with those observed in the experiments performed by \citet{kim2009,kim2009Three,cevheri2014,yuan2016, xuan2018,yee2018}.

It is of interest to determine the underlying interactions which cause the shift in the equilibrium positions. When a rigid particle is suspended in shear flow, it gives rise to a stresslet disturbance. Upon the application of an external electric field, an additional source-dipole disturbance emerges due to the electrokinetic slip. 
These disturbance fields (stresslet and slip-originated source-dipole) upon mutual interaction among themselves; upon interaction with the background flow (shear $ \beta $ and curvature $ \gamma $) and walls, contribute to the electrophoresis combined inertial lift. 
To interpret the underlying mechanisms which cause the shift in migration, we dissect the forces into various constituents via an order-of-magnitude analysis. We then evaluate and explain these constituents separately (as in \S \ref{sec:6}.2). 

The order of background velocity in terms of outer coordinates is given by:

\floatsetup[figure]{style=plain,subcapbesideposition=top}

\begin{equation}
\tilde{\IB{V}}_\infty ^{\left( 0 \right)}\sim O\left( {{1 \mathord{\left/
			{\vphantom {1 \kappa }} \right.
			\kern-\nulldelimiterspace} \kappa }} \right)\beta  + O\left( {{1 \mathord{\left/
			{\vphantom {1 {{\kappa ^2}}}} \right.
			\kern-\nulldelimiterspace} {{\kappa ^2}}}} \right)\gamma  + O\left( 1 \right)\gamma +O(1)\, H\!a Z_{p},
\label{7.1}
\end{equation}
the order of disturbance velocity (stresslet and source-dipole) field is:
\begin{equation}
{\tilde{\IB{v}}^{\left( 0 \right)}} \sim O\left( {{\kappa ^2}} \right)\beta \left( {1 + {W_D}} \right) + O\left( {{\kappa ^3}} \right)Ha{Z_p}\left( {1 + {W_B}} \right),
\label{7.2}
\end{equation}
and the order of test velocity field is: 
\begin{equation}
 \tilde{\IB{u}}^{t} \sim O(\kappa)(1+W_{A}^{t}) .
\label{7.3}
\end{equation}
Substituting the above expressions in the lift force (\ref{4.12}), we obtain seven components which contribute to the inertial force experienced by an electrophoretic particle. These arise from various interactions in the system (see table \ref{table2}). The first two components (SWF and SCF) belong to the lift associated with an inert particle suspended in Poiseuille flow and were discussed in \S \ref{sec:6}.2. The remaining five components (table \ref{table2}: \#3-7), arising from the introduction of electrophoresis, can be lumped into what we term as ‘phoretic lift’ (as this lift is general for any phoretic transport such as thermophoresis or diffusiophoresis). Figure \ref{fig:6}(b) compares the phoretic lift on a positive $ Z_{p} $ particle with the stresslet-wall force (SWF) and stresslet-curvature force (SCF). A positive slope at the centerline depicts that the phoretic lift acts in the direction of SCF and contributes to pushing the leading particle towards the walls. This results in an outward shift of equilibrium positions i.e. a leading particle finds equilibrium nearer to the walls. 

The analytical expression derived in \S \ref{sec:4}.1 $ (-47 \pi \beta H\!a Z_{p} \Rey_{p}/20) $ is plotted as a dashed line in figure \ref{fig:6}(b). Since it is proportional to the background shear, the trend remains linear throughout the channel width. 
It can also be seen from figure \ref{fig:6}(b) that the phoretic lift (which incorporates the wall correction) agrees with the analytical expression away from the walls. The increased viscous resistance near the walls reduces the phoretic lift, and causes a deviation from the linear analytical expression.
The effect of wall on the phoretic lift is weak because it is short-ranged, as it is caused by rapidly decaying source-dipole field ($ O(1/r^{3}) $). On the other hand, the stresslet-wall force (SWF) and stresslet-curvature force (SCF) are characterized by a less rapidly decaying stresslet field ($ O(1/r^{2}) $). Therefore, SWF and SCF are influenced sufficiently far away from the walls.

\setlength{\dashlinedash}{2pt}
\begin{table}
	\renewcommand{\arraystretch}{0.6}
	\small
	\begin{center}
		\def~{\hphantom{0}}
		\begin{tabular}{l c c c}\\ \hline\\[1pt]
			\#                                                      & Component                            & Integrand        & Migration towards \\[5pt] \hline \\[3pt]
			\begin{tabular}[t]{c} 1 \end{tabular}       & \begin{tabular}[t]{c} Stresslet-Wall\\ (SWF) \end{tabular}      & \begin{tabular}[t]{c} $ O(1)\beta\beta(1+W_{D}) $ \end{tabular}     & \begin{tabular}[t]{l} Center \end{tabular} \\ \\
			
			\begin{tabular}[t]{c} 2 \end{tabular}       & \begin{tabular}[t]{c} Stresslet-Curvature \\(SCF) \end{tabular}      & \begin{tabular}[t]{c} $ O(1/\kappa)\gamma\beta(1+W_{D}) $ \end{tabular}     & \begin{tabular}[t]{l} Walls \end{tabular} \\ 
			
			\hdashline \\
			
			\begin{tabular}[t]{c} 3 \end{tabular}       & \begin{tabular}[t]{c} Stresslet-Phoretic \\(SPF\textsubscript{E}) \end{tabular}      & \begin{tabular}[t]{c} $ O(\kappa^{5})\beta H\!a Z_{p} (1+W_{B})(1+W_{D}) $ \end{tabular}     & \begin{tabular}[t]{l} for $ H\!a Z_{p}>0 $, walls\\ for $ H\!a Z_{p}<0 $, center \end{tabular} \\ \\
			
			\begin{tabular}[t]{c} 4 \end{tabular}       & \begin{tabular}[t]{c} Phoretic-Shear\\ (PSF\textsubscript{E}) \end{tabular}      & \begin{tabular}[t]{c} $ O(\kappa)\beta H\!a Z_{p}(1+W_{B}) $ \end{tabular}     & \begin{tabular}[t]{l} for $ H\!a Z_{p}>0 $, walls\\ for $ H\!a Z_{p}<0 $, center \end{tabular} \\ \\
			
			\begin{tabular}[t]{c} 5 \end{tabular}       & \begin{tabular}[t]{c} Stresslet-Lag/Lead\\ (SLF\textsubscript{E}) \end{tabular}      & \begin{tabular}[t]{c} $ O(\kappa) H\!a Z_{p}\beta(1+W_{D}) $ \end{tabular}     & \begin{tabular}[t]{l} for $ H\!a Z_{p}>0 $, walls\\ for $ H\!a Z_{p}<0 $, center \end{tabular} \\ \\
			
			\begin{tabular}[t]{c} 6 \end{tabular}       & \begin{tabular}[t]{c} Phoretic-Curvature\\ (PCF\textsubscript{E}) \end{tabular}      & \begin{tabular}[t]{c} $ O(1)\gamma H\!a Z_{p}(1+W_{B}) $ \end{tabular}     & \begin{tabular}[t]{l} for $ H\!a Z_{p}>0 $, center\\ for $ H\!a Z_{p}<0 $, walls \end{tabular} \\ \\
			
			\begin{tabular}[t]{c} 7 \end{tabular}       & \begin{tabular}[t]{c} Phoretic-Wall\\ (PWF\textsubscript{E}) \end{tabular}      & \begin{tabular}[t]{c} $ O(\kappa^{2}) (H\!a Z_{p})^{2}(1+W_{B}) $ \end{tabular}     & \begin{tabular}[t]{l} Center \end{tabular} 
			\\ \hline
		\end{tabular}
		\caption{Components which constitute the electrophoresis combined inertial lift force. The forces arising due to electrophoresis are subscripted by ‘E' because the external agency is electric field. The first two components are the same as described in figure \ref{fig:5}.}
		\label{table2}
	\end{center}
\end{table}
\normalsize

\begin{figure}%
	\centering
	\sidesubfloat[]{{\includegraphics[scale=0.38]{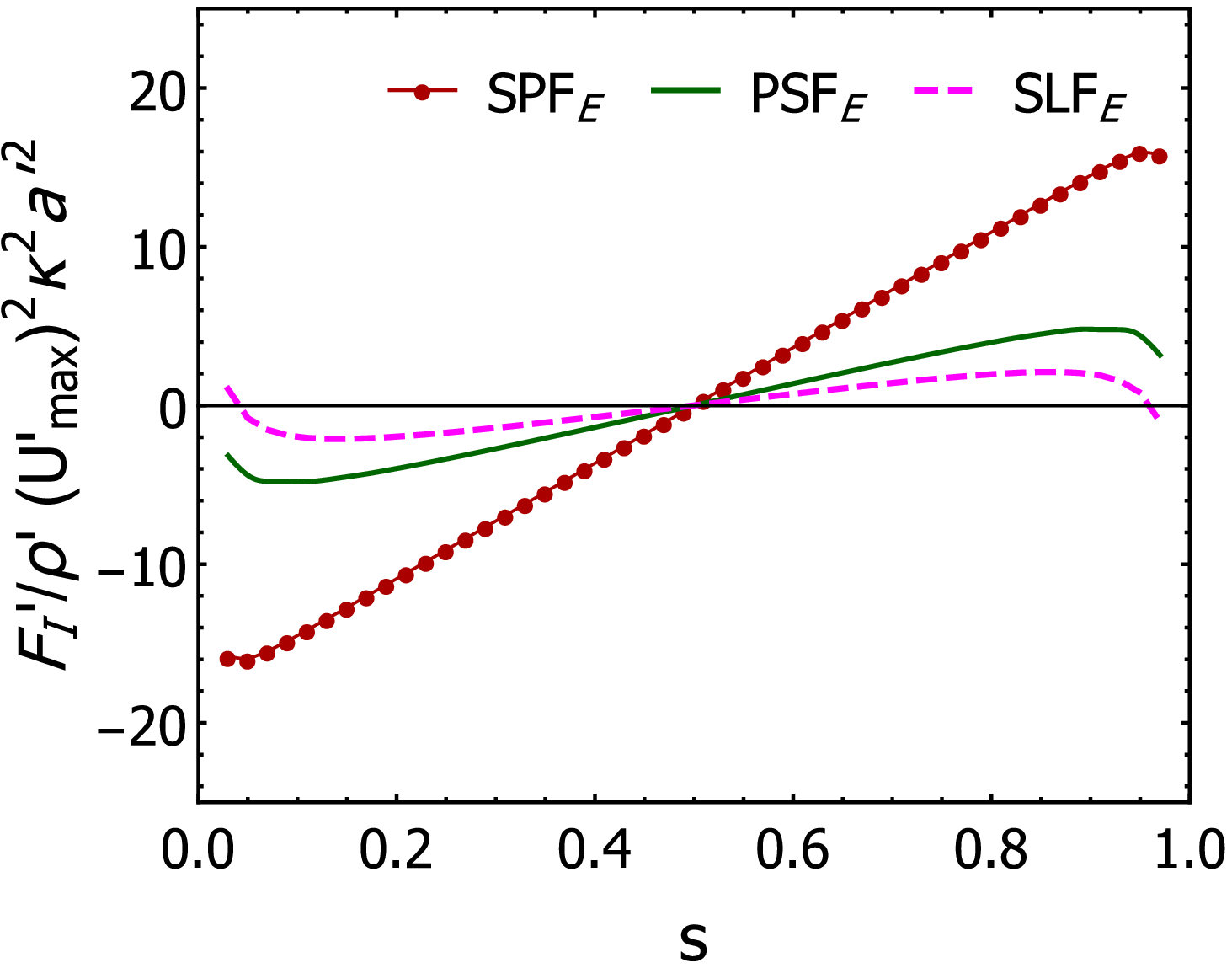} }}%
	\qquad
	\sidesubfloat[]{{\includegraphics[scale=0.296]{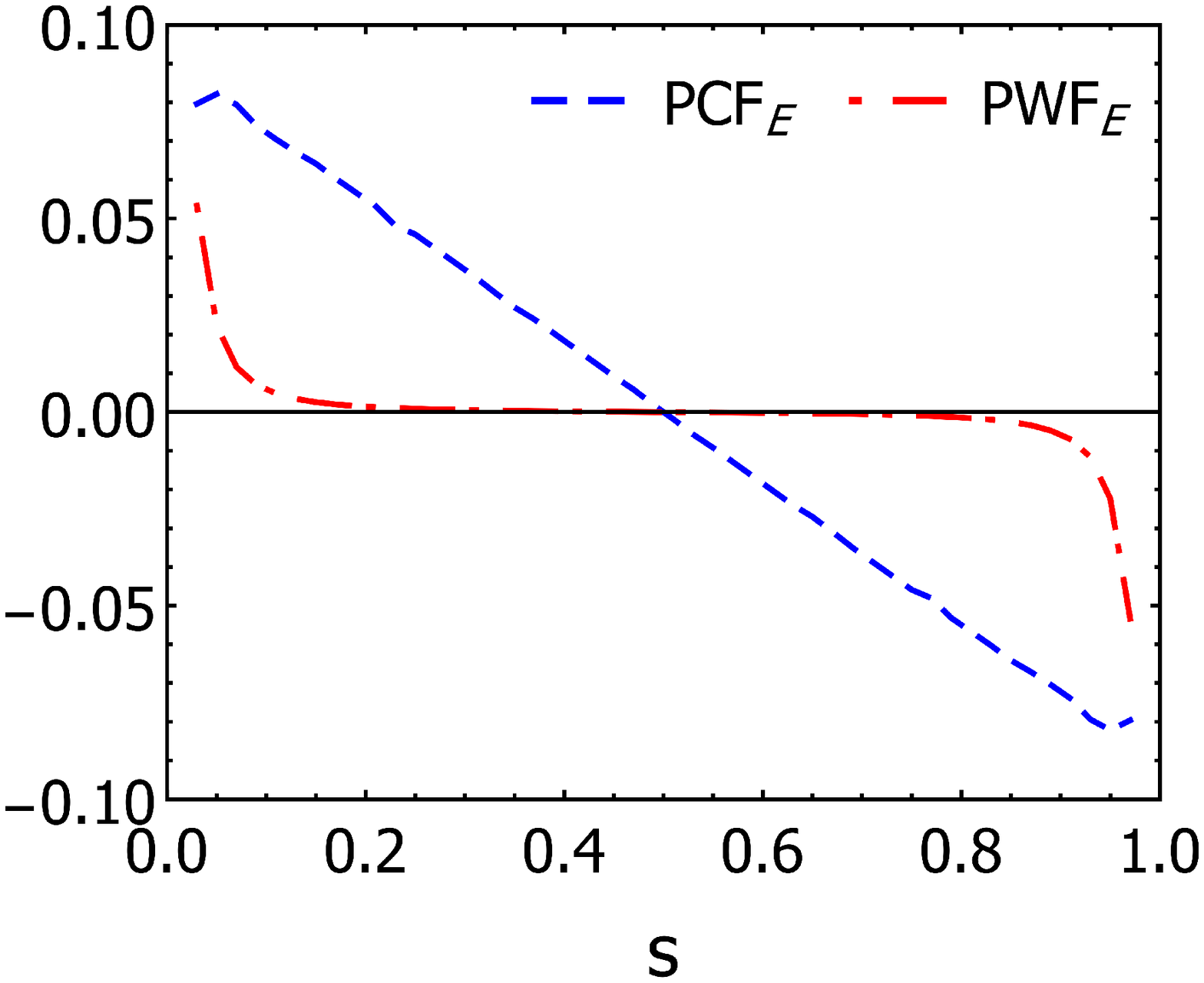} }}%
	\caption{Variation of (a) major and (b) minor components of the phoretic-lift, for a leading particle. See table \ref{table2} for the details of abbreviations. Parameters:  $ \zeta_{p}'=47\: \mbox{mV,\ } U_{max}'=2.25\: \mbox{mm/s,\ } a'=10\: \mu \mbox{m,\ } l'=1 \mbox{\:mm,\ } \varepsilon'=80\,\varepsilon_{0}' \mbox{,\ } \mu'=10^{-3} \mbox{\:Pa\,s,\ }  $$ E_{\infty}'= $70 V$ / $cm.}%
	\label{fig:7}%
\end{figure}

The phoretic lift force is composed of five components listed in table \ref{table2} (\#3-7). Figure \ref{fig:7} illustrates their variation with respect to lateral position. Major components which govern the phoretic-lift are plotted in figure \ref{fig:7}(a). The disturbance-disturbance interaction term $ (\IB{v}^{(0)} \bcdot \bnabla \IB{v}^{(0)}) $ (\#3 in table\ref{table2}) contributes primarily to the phoretic-lift. The proportionality to $ \beta H\!a Z_{p} $ suggests that this must arise from the interaction between stresslet and source-dipole disturbance, and is therefore classified as stresslet-phoretic force (SPF\textsubscript{E}). This SPF\textsubscript{E} acts to push the leading particle towards the walls. These interactions are short ranged, and therefore the associated force decays (due to viscous resistance) only when the particle is close to the walls.
The other two significant forces arise from the interaction of disturbance field and the background flow ($ (\IB{V}_{\infty}^{(0)} \bcdot \bnabla \IB{v}^{(0)}) $ and $ (\IB{v}^{(0)} \bcdot \bnabla \IB{V}_{\infty}^{(0)}) $). The interaction of phoretic disturbance with the background shear results in a phoretic-shear force (PSF\textsubscript{E}).  A stresslet-lead/lag force (SLF\textsubscript{E}) occurs when the quadrupolar symmetry of pressure around the particle (characteristic of stresslet disturbance) is distorted due to an externally imposed lead/lag. From figure \ref{fig:7}(a) we conclude that when a particle leads the flow (i.e. $ H\!a Z_{p}>0 $), all the three components: SPF\textsubscript{E}, PSF\textsubscript{E}, and SLF\textsubscript{E} push it towards the walls, while the reverse is true for a lagging particle.

\floatsetup[figure]{style=plain,subcapbesideposition=top}

Other minor components of the phoretic-lift are shown in figure \ref{fig:7}(b) (compare the y-axis scale). Contrary to the background shear, the curvature interacts with the source-dipole disturbance to produce lift (phoretic-curvature force) in the opposite direction but at a much lower magnitude. The phoretic-wall force (PWF\textsubscript{E}) occurs due to the interaction of rapidly decaying phoretic disturbance with the walls (analogous to stresslet-wall force). This weak PWF\textsubscript{E} would exist in isolation for inertial migration in a pure electrokinetic flow (i.e. absence of external pressure drop). Moreover, PWF\textsubscript{E} always acts towards the centerline irrespective of the sign of the electric field or zeta potential, as it is proportional to $ (H\!a Z_{p})^{2} $.

From the results, we observe that, in Poiseuille flow, the effect of electrophoresis on the inertial migration (phoretic-lift) is qualitatively similar to that of a non-neutrally buoyant particle \citep{saffman1965} i.e. a particle leading the flow, migrates towards the region of high shear and vice-versa for a lagging particle.
	Therefore, it is of interest to find the fundamental differences between the two systems. Towards this, in the next section, we investigate the influence of a force driven phenomenon (buoyancy) on the inertial migration. We then differentiate it from the slip-driven phenomenon (electrophoresis) in terms of its influence on the inertial migration.

%%%%%%%%%%%%%%%%%%%%%%%%%%%%%%%%%%%%%%%%%%%%%%%%%%%%%%%
%%%%%%%%%%%%%%%%%%%%%%%%%%%%%%
\section{Comparisons with buoyancy combined inertial lift}\label{sec:8}
In the previous section, we analyzed the influence on the inertial lift due to an external field, which makes the particle lag or lead the flow via a ‘force-free mechanism’. We now focus our attention on the inertial lift experienced by a non-neutrally buoyant inert particle suspended in a Poiseuille flow. The gravitational field acts parallel to the flow and makes the particle lag/lead via a ‘forced mechanism’. The associated disturbance profile is fundamentally different from the force-free scenario (see figure \ref{fig:1}). Therefore, we focus on the influence of these differences on particle migration.

Following the procedure described in \S \ref{sec:3}, the translational and rotational velocities are found by balancing the hydrodynamic drag with buoyancy and rendering the particle torque free. This results in: 
\begin{equation}
U_{s\,x}^{(0)} \approx \frac{{(\alpha + \gamma/3)({1 + \kappa {W_A}}) - {{10{\kappa ^2}\beta {W_D}}/9 \; + \; {\cal B}}}}{( {1 + \kappa {W_A}} )},
\label{8.1}
\end{equation}
\begin{equation}
\Omega_{s\,y}^{(0)} \approx  \frac{\beta ( {1 + \kappa {W_A}}) - (10{{\cal X}_D}{\kappa ^3}\beta /3) + 3{\cal B}{\cal{X}}_{A} \kappa^2}{2(1+\kappa W_{A})} .
\label{8.2}
\end{equation}
Here, $ {\cal B} $ is the buoyancy number defined as the ratio of buoyancy to viscous drag force: 
\begin{equation}
{\cal B} = \frac{{\Delta \rho' \left( {{{4\pi {a'^3}} \mathord{\left/
					{\vphantom {{4\pi {a'^3}} 3}} \right.
					\kern-\nulldelimiterspace} 3}} \right)g'}}{{6\pi \mu' a'{U_{max}'} \kappa}}.
\label{8.3}
\end{equation}
Here, $ \Delta \rho $ is the density difference between the particle and fluid medium and $ g $ is the acceleration due to gravity.

\begin{figure}
	\centerline{\includegraphics[scale=0.42]{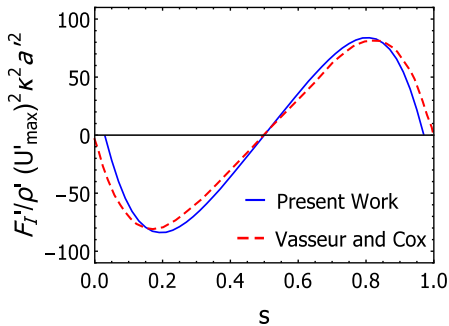}}% Images in 100% size
	\caption{Lift force on a non-neutrally buoyant particle in a Poiseuille flow and comparison with \citet{vasseur1976}. Parameters: $ a' $=10 $ \mu $m, $ l' $=1 mm, $ U_{max}' $=2.25 mm/s, $ \mu' $=10\textsuperscript{-3} Pa s and $ \Delta \rho' $=100 kg/m\textsuperscript{3}.}
	\label{fig:8}
\end{figure}

In \S \ref{sec:7}, we analyzed the effect of electrophoresis on the inertial migration for $Ha{Z_p} \sim O\left( 1 \right)$. The corresponding electrophoretic velocity was of the order of characteristic velocity i.e. $ O(\IB{U}_{E}')  \sim O(\kappa U'_{max}) $. For an equitable comparison, we choose the buoyancy number ($ {\cal B} $) to be such that the lag/lead produced by the density difference is of the same order as that in the electrophoretic system.
$ \mathcal{B} \sim O(1) $ corresponds to the settling velocity of the sphere to be of the order of characteristic velocity: $ U_{set}'\sim \kappa U_{max}' $. Here $ U_{set}' $ is the settling velocity corresponding to $ {\cal B} $.
Figure \ref{fig:8} depicts the lift force corresponding to $ {\cal B}=1 $. The curve shows that, for the flow aligned in the direction of gravity, a heavy particle ($ {\cal B} > 0 $) migrates towards the walls.
Our result agrees well with that of \citet{vasseur1976} who addressed a similar regime of buoyancy: $ \kappa^{2}U_{max}' \ll U_{set}' \ll U_{max}' $.  A slight mismatch arises near the walls because their analysis assumes that the particle settles with a constant velocity for all ‘$ s $’ while we take wall resistance into account which reduces the settling velocity significantly near the walls.

%\citet{feng1994} have reasoned based on their numerical results, that the interaction of the stokeslet disturbance (generated by the external gravitational fields) with the background shear results in a shift of equilibrium positions. The approach employed in this work allows us to explain these observations by explicitly observing various components which contribute to the buoyancy combined inertial lift. 
\setlength{\dashlinedash}{2pt}
\begin{table}
	\renewcommand{\arraystretch}{0.6}
	\small
	\begin{center}
		\def~{\hphantom{0}}
		\begin{tabular}{l c c c}\\ \hline\\[1pt]
			\#                                                    & Component                            & Integrand        & Migration towards \\[5pt]\hline \\[3pt]
			
			\begin{tabular}[t]{c} 1 \end{tabular}       & \begin{tabular}[t]{c} Stokeslet-Shear\\ (SSF\textsubscript{G}) \end{tabular}      & \begin{tabular}[t]{c} $ O(\kappa)\beta {\cal B} (1+W_{A}) $ \end{tabular}     & \begin{tabular}[t]{l} for $ {\cal B}>0 $, walls\\ for $ {\cal B}<0 $, center \end{tabular} \\ \\
			
			\begin{tabular}[t]{c} 2 \end{tabular}       & \begin{tabular}[t]{c} Stresslet-Lag/Lead\\ (SLF\textsubscript{G}) \end{tabular}      & \begin{tabular}[t]{c} $ O(\kappa) {\cal B} \beta(1+W_{D}) $ \end{tabular}     & \begin{tabular}[t]{l} for $ {\cal B}>0 $, walls\\ for $ {\cal B}<0 $, center \end{tabular} \\ \\
			
			\begin{tabular}[t]{c} 3 \end{tabular}       & \begin{tabular}[t]{c} Stokeslet-Curvature\\ (SCF\textsubscript{G}) \end{tabular}      & \begin{tabular}[t]{c} $ O(1)\gamma {\cal B} (1+W_{A}) $ \end{tabular}     & \begin{tabular}[t]{l} for $ {\cal B}>0 $, center\\ for $ {\cal B}<0 $, walls \end{tabular} \\ \\
			
			\begin{tabular}[t]{c} 4 \end{tabular}       & \begin{tabular}[t]{c} Stokeslet-Wall\\ (SWF\textsubscript{G}) \end{tabular}      & \begin{tabular}[t]{c} $ O(\kappa^{2}) {\cal B}^{2} (1+W_{A}) $ \end{tabular}     & \begin{tabular}[t]{l} Center \end{tabular} 
			\\ \hline
		\end{tabular}
		\caption{Lift force components which comprise the additional lift force occurring due to a non-neutrally buoyant particle. The subscript ‘G' denotes that the forces arise from external gravitational field.}
		\label{table3}
	\end{center}
\end{table}
\normalsize

\floatsetup[figure]{style=plain,subcapbesideposition=top}
\begin{figure}%
	\centering
	\sidesubfloat[]{{\includegraphics[scale=0.290]{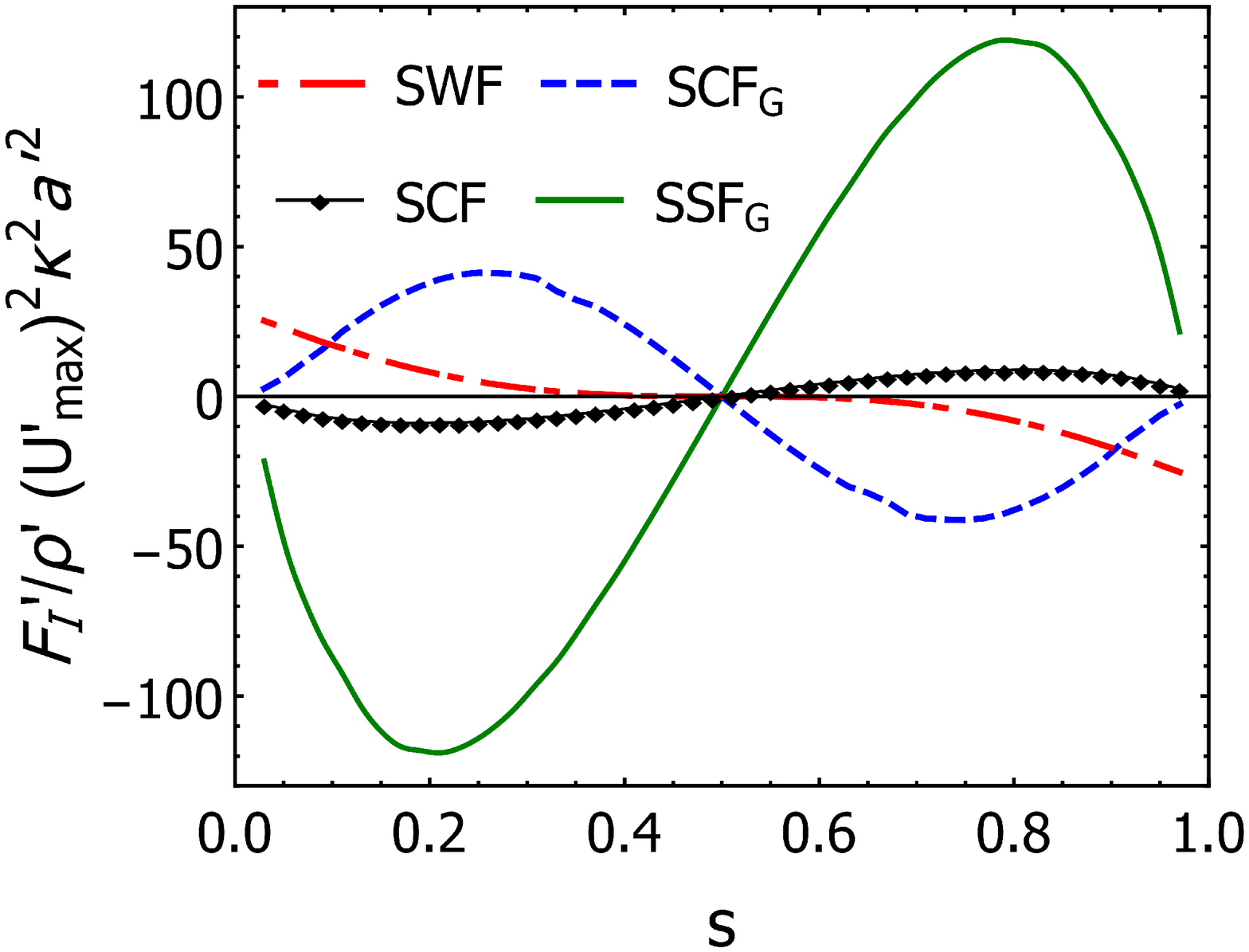} }}%
	\qquad
	\sidesubfloat[]{{\includegraphics[scale=0.41]{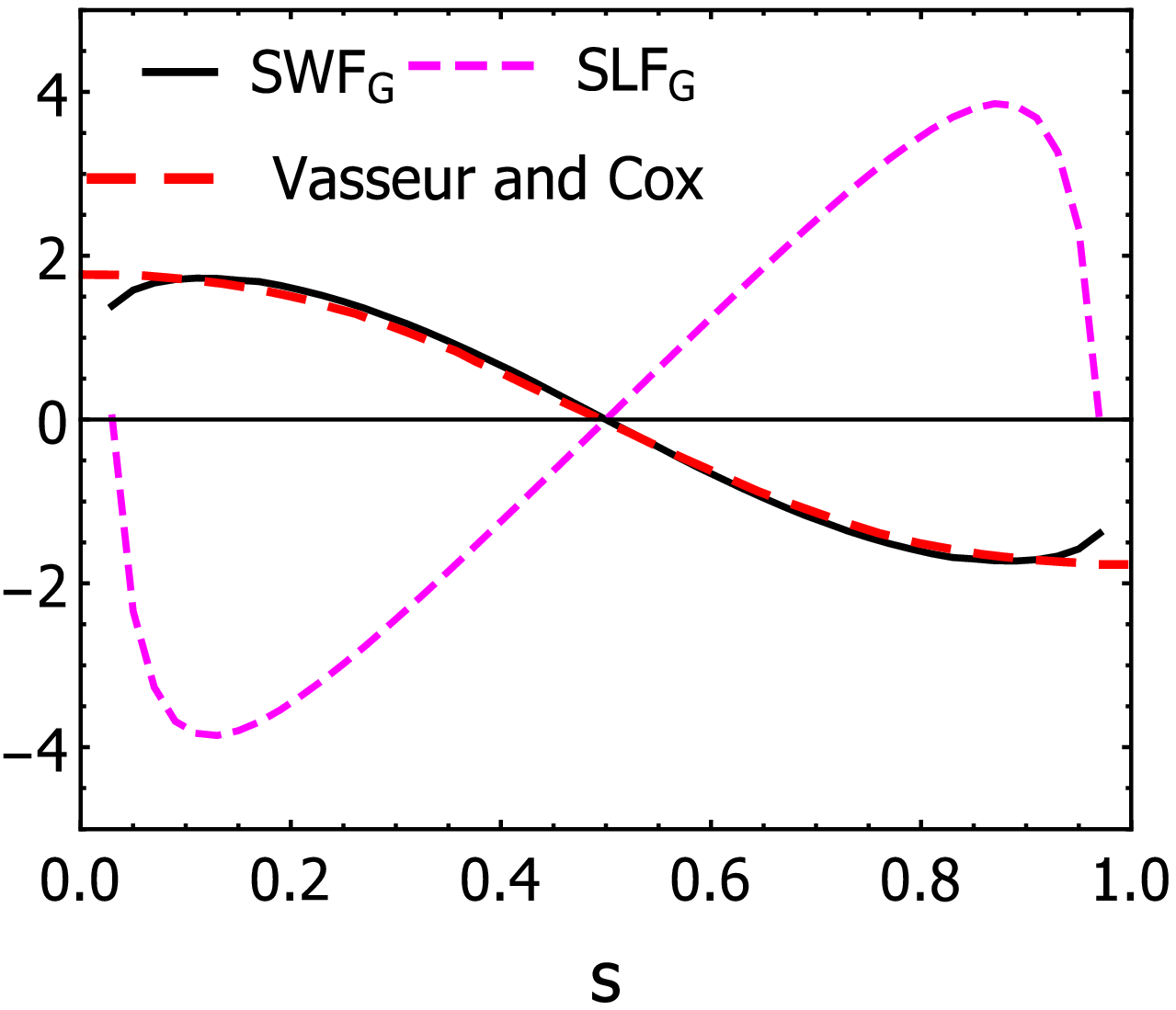} }}%
	\caption{Variation of (a) major components (SSF\textsubscript{G} and SCF\textsubscript{G}) and (b) minor components (SWF\textsubscript{G} and SLF\textsubscript{G}) of buoyancy combined inertial force as a function of distance from the bottom wall (see table \ref{table3}). Parameters: $ a' $=10 $ \mu $m, $ l' $=1 mm, $ U_{max}' $=2.25 mm/s, $ \mu' $=10\textsuperscript{-3} Pa s and $ \Delta \rho' $=100 kg/m\textsuperscript{3}.}%
	\label{fig:9}%
\end{figure}

As in \S \ref{sec:7}, we analyze the order-of-magnitude of various components of the inertial lift for a non-neutrally buoyant particle. We obtain four additional components apart from stresslet-wall force (SWF) and stresslet-curvature force SCF (see \S \ref{sec:6}.2). Table \ref{table3} illustrates these additional components of buoyancy combined inertial lift, which arise from the various interactions. Contrary to the electrophoretic case, the contribution emerging due to the disturbance-disturbance interactions is found to be negligible in this case and hence are not discussed here. 
We also find that the expressions for the components contributing to the lift, emerging due to buoyancy (table \ref{table3}, \#1-4), are analogous to those arising from electrophoresis (c.f. table \ref{table2}, \#4-7). 
Also, the components of the two cases show similar effects on migration. For instance, both phoretic-shear force and stokeslet–shear force direct the leading particle to focus at walls, while phoretic-curvature and stokeslet-curvature force direct the leading particle towards the center.

%\begin{figure}
%	\centerline{\includegraphics[scale=0.6]{NNB_compo}}% Images in 100% size
%	\caption{Variation of (a) leading constituents of buoyancy combined inertial force as a function of distance from the bottom wall and of (b) minor components. Parameters for: $ a $=10 $ \mu $m, $ l $=1 mm, $ U_{max}{'} $=2.25 mm/s, $ \mu $=10\textsuperscript{-3} Pa s and $ \Delta \rho $=100 kg/m\textsuperscript{3}.}
%	\label{fig:9}
%\end{figure}

In figure \ref{fig:9}(a), we show that stokeslet-shear (SSF\textsubscript{G}) and stokeslet-curvature forces (SCF\textsubscript{G}) dominate over the SWF and SCF (emerging from stresslet, discussed in \S \ref{sec:6}.2). SSF\textsubscript{G} and SCF\textsubscript{G} primarily govern the migration for a non-neutrally buoyant particle; unlike the electrophoretic case where the phoretic-lift was found to be comparable to SWF and SCF (see figure \ref{fig:6}(b)). We conclude, on the basis of equitable lead/lag, that the modification to inertial lift arising due to density difference is much greater than that caused due to electrophoresis. Furthermore, we observe that the components involving interaction with the background curvature (PCF\textsubscript{E} and SCF\textsubscript{G}), oppose the interactions involving the background shear (PSF\textsubscript{E} and SSF\textsubscript{G}), irrespective of the external driving mechanism (such as gravity or electric field).

If the ratio of settling velocity to the background flow is asymptotically large (i.e. $ U_{set}'\gg U_{max}' $), only the stokeslet-wall force (SWF\textsubscript{G}) exists. This corresponds to a sedimentation of a particle in a quiescent fluid ($ \beta \equiv 0 \mbox{ and\ } \gamma \equiv 0 $). The particle would focus at the centerline for both positive and negative density differences (as SWF\textsubscript{G} is proportional to $ {\cal B}^{2} $). In figure \ref{fig:9} (b), we find that the SWF\textsubscript{G} matches well with the inertial lift reported by Vasseur and Cox (1976) for a sedimenting particle (in quiescent fluid). The deviation near the walls is attributed to the inclusion of first order wall correction in our approach (see \ref{8.1}). Whereas \citet{vasseur1976} assumed a constant settling velocity irrespective of the particle position between the walls.

\section{Comparisons with the decoupled electric lift}\label{sec:9}

In \S \ref{sec:3}.1, we showed that the electrical force $ \IB{F}_{M} $ (arising from the Maxwell stresses)  contributes to the lateral migration of the particle (see eq. \ref{FM}).
This force pushes the particle away from the walls, irrespective of the particle zeta potential. On the other hand, the inertial lift on an electrophoretic particle depends on the zeta potential (see figure \ref{fig:6}(a)).
Therefore, the equilibrium positions of an electrophoretic particle will be determined by the relative magnitudes of the electric lift (EL) and the electrophoresis combined inertial lift (EIL).
Here we compare the contributions arising from these two effects.

\begin{figure}%
	\centering
	\sidesubfloat[]{{\includegraphics[scale=0.4]{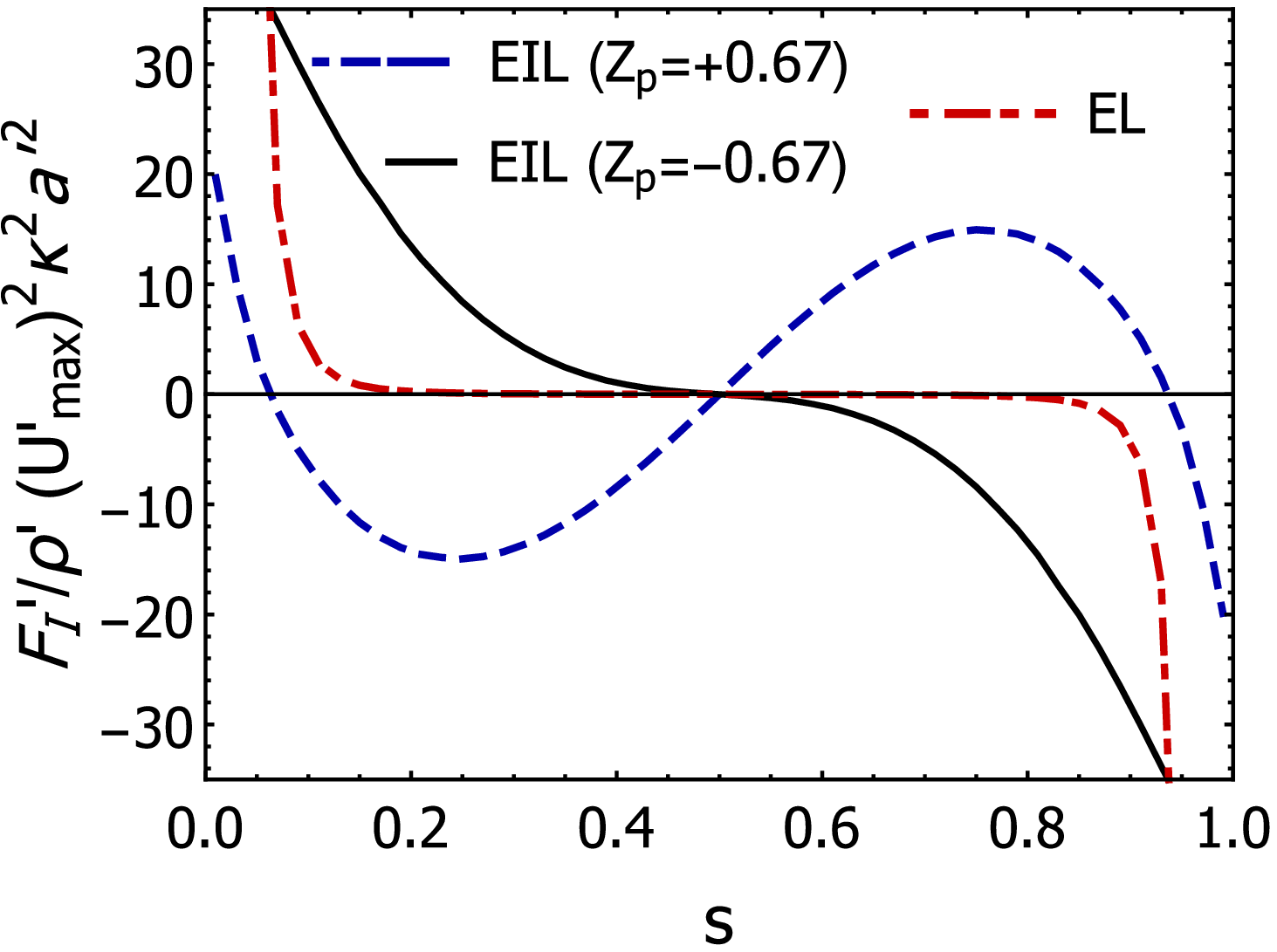} }}%
	\qquad
	\sidesubfloat[]{{\includegraphics[scale=0.39]{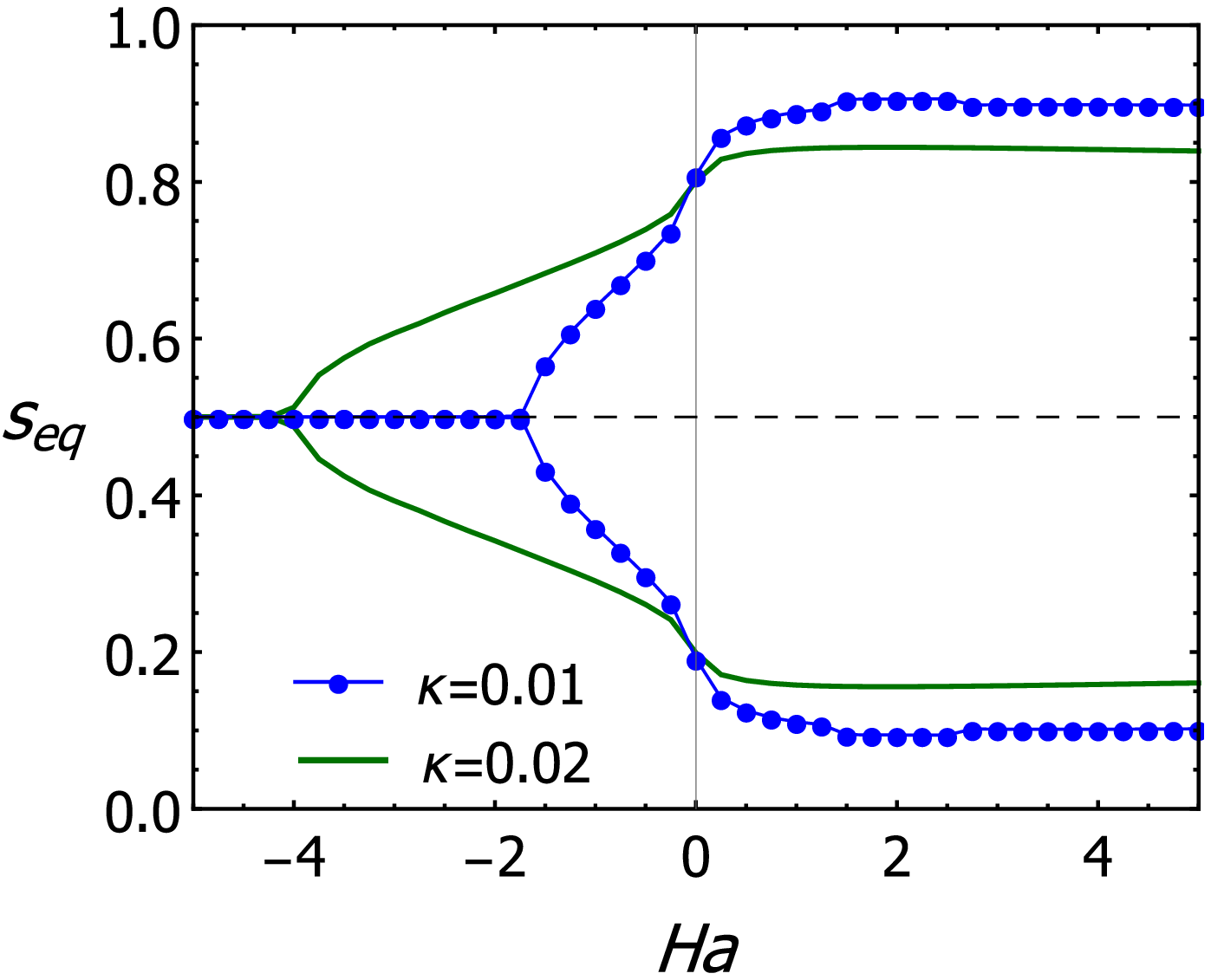} }}%
	\caption{(a) Comparison of electrophoresis combined inertial lift (EIL) with the electric lift (EL) (\ref{FM}) for $ \zeta_{p}' $=47 mV and $ \zeta_{p}' $=-47 mV. Parameters: $ H\!a $ =+1.25; $ E_{\infty}' $=70 V/cm, $ U_{max}' $=2.25 mm/s, $ a' $=10 $ \mu $m, $ l' $=1mm, $ \varepsilon'=80 \varepsilon_{0}' $, $ \mu' $=10\textsuperscript{-3} Pa s, $ \rho' = 10^{3}\, $kg/m\textsuperscript{3}. (b) Effect of $ H\!a  $ and size ratio ($ \kappa=0.01 $ and $ \kappa=0.02 $) on the equilibrium positions ($ \zeta_{p}' $=47 mV). A negative $ H\!a $ implies that the electric field and the flow are in the opposite directions.}%
	\label{fig:10}%
\end{figure}

In figure \ref{fig:10}(a), we show the variation of the two lifts with lateral position between the walls, where the electric field is applied in the direction of the flow ($ H\!a > 0 $).
For $ \kappa \ll 1 $, we find that the wall-repulsive EL decays rapidly away from the wall and becomes negligible in the bulk of the channel. 
The migration of an electrophoretic particle is therefore determined by EIL in the bulk of the channel.

Figure \ref{fig:10}(b) shows the variation of equilibrium positions ($ s_{eq} $) with the strength of the applied electric field ($ H\!a $). Here, the equilibrium positions are determined by the total lateral lift on the particle (i.e EL+EIL).
The solid lines depict stable equilibrium positions, whereas the dashed line represents an unstable equilibrium. The curves resemble a classic supercritical pitchfork bifurcation \citep{strogatz}. 
As we increase the electric field strength in the negative $ x-$direction, for a fixed zeta potential ($ \zeta_{p}' $=+47 mV) and particle to channel size ratio ($ \kappa=0.01 $), a critical $ H\!a $ exists ($ \approx -1.8 $), below which the particle focuses only at the centerline.
At this point, the center equilibrium point undergoes a transition from an unstable to stable equilibrium because the phoretic-lift is sufficiently strong to overcome the stresslet-curvature force (SCF).

\floatsetup[figure]{style=plain,subcapbesideposition=top}

As the electric field strength is increased in the positive $ x- $direction, the equilibrium positions shift towards the walls. 
We can also observe from figure \ref{fig:10}(b) that a strong electric lift prevents the particle from focusing near the walls. This effect is prominent for a larger particle ($\kappa$=0.02).
%, as the electric lift is proportional to $ a^{6} $ ($ =a^{2}\kappa^{4} $), as compared to the $ a^{4} $ dependency of the inertial forces. 
The analysis in this section helps establish that the separation of particles can be carried out based on size, surface charge, and external electric field.

%%%%%%%%%%%%%%%%%%%%%%%%%%%%%%%%%%%%%%%%%%%%%%%%%%%%%%%%%%%%%%%%%%%%%%%%%%%%%%%%%%%%%%%%

\section{Conclusions}\label{sec:10}

 Motivated by the recent experimental observations \citep{kim2009,cevheri2014,yuan2016,xuan2018}, we analyze the inertial lift on an electrophoretic particle suspended in a combined electro-osmotic Poiseuille flow by using the framework of \citet{ho1974} . 
We use the reciprocal theorem in conjunction with perturbation expansion (in particle Reynolds number) to obtain the lift arising from weak inertia. The Stokes velocity fields are found by using the method of reflections; odd reflections are obtained using Lamb's solution \citep{lamb} and even reflections are obtained using Faxen transform \citep{faxen1922}. By neglecting the wall effects, we obtain an analytical expression for the `phoretic-lift' which captures the influence of electrophoresis on inertial migration. Later we include the wall effects at the leading order.  We find that our results agree qualitatively with the experimental observations;  a leading particle migrates towards the regions of low background velocity (i.e. walls) while the reverse occurs for a lagging particle.

The chosen framework allows us to perform an order-of-magnitude analysis and obtain various components which contribute to the inertial lift. We classify these components on the basis of the interactions which cause them. 
Our results reveal that the phoretic-lift arises primarily due to short-ranged interactions between electrokinetic slip-driven source-dipole field and the stresslet field (generated due to particle resistance to strain in the background Poiseuille flow). 
We observe that, the effect of electrophoresis on the inertial migration (phoretic-lift) is qualitatively similar to that of non-neutrally buoyant particle \citep{saffman1965}.
Therefore, to differentiate the slip-driven phenomenon (electrophoresis) from a force-driven phenomenon (buoyancy) in terms of their influence on the inertial migration, we further study the migration of a non-neutrally buoyant particle in Poiseuille flow.
%Motivated by the fact that a slip-driven motion is fundamentally different from force-driven motion \citep{anderson1989}, we draw comparisons between electrophoresis and the gravitational effects in terms of their influence on the migration.
% Towards this, we further study the case of a non-neutrally buoyant particle, where the flow is aligned parallel to the gravitational field. 
The buoyancy driven motion of the particle generates a Stokeslet field at the leading order. We find that the interaction of this field with the background shear primarily influences the inertial migration. 
Contrary to electrophoresis, this interaction is long-ranged and the corresponding lift is much stronger than the phoretic-lift.

Finally, we show that the particles can be focused near the center (or at the walls) if a sufficiently high electric field is applied in a suitable direction. 
A decoupled wall-repulsive electric lift (derived in this work as an extension to \citet{yariv2006}) may prevent the particle from focusing near the walls and hinder the particle separation.
For asymptotically small particles ($ \kappa \ll 1 $), this electric lift decays sharply away from the walls, and does not affect the equilibrium positions in the bulk of the channel.

Although this work is limited to weak inertia, we believe that these findings provide a first step towards understanding the influence of electric fields on inertial migration of particles, which provides a good starting point for further research. For instance, the problem becomes challenging for higher Reynolds numbers as the streaming potential (arising from the convection of ions in the double layer) might contribute significantly to the inertial lift \citep{schnitzer2012}. Furthermore, for high particle zeta potentials, surface conduction effects might significantly alter the electrokinetic-slip and cause substantial deviation from the inertial lift. These questions may constitute the objective for future studies.
\\ \\

First author A.C is indebted to Dr. Anubhab Roy of the Applied Mechanics Department at Indian Institute of Technology Madras for his lectures on Microhydrodynamics and fruitful discussions. The authors are also thankful to Dr. Sumesh Thampi of the Chemical Engineering Department at Indian Institute of Technology Madras for his feedback. A.C. is also grateful to MHRD, Gov. of India, for financial support.

%\newpage

%%%%%%%%%%%%%%%%%%%%%%%%%%%%%%%%%%%%%%%%%%%%%%%%%%%%%%%%%%%%%%%%%%%%%%%%%%%%%%%%%%%%%%%%
\appendix
\section{}\label{appA} 
\subsection{Justification of neglect of translational and angular acceleration}
In this appendix, we justify the force-free and torque-free assumption in (\ref{D-F}) and (\ref{D-L}). We choose the time scale to be $ t_{conv}' $, and obtain the dimensionless force and torque balance as:
\begin{eqnarray}
&\IB{F} = \int\limits_{{S_p}} {\IB{n} \bcdot {\IB{\sigma} _H}\mathrm{d}S}   +  4\pi H\!a\int\limits_{{S_p}} {\IB{n} \bcdot {\IB{\sigma} _M}\mathrm{d}S}  = \frac{4\pi}{3} \Rey_{p} \frac{\partial{\IB{U}_{s}}}{\partial {t}} \nonumber \\
&\IB{L} = \int\limits_{{S_p}} {\IB{n} \times \left( {\IB{n} \bcdot {\IB{\sigma} _H}} \right)\mathrm{d}S}   +  4\pi H\!a\int\limits_{{S_p}} {\IB{n} \times \left( {\IB{n} \bcdot {\IB{\sigma} _M}} \right)\mathrm{d}S}  = \frac{4\pi}{3} \Rey_{p} \frac{\partial{\IB{\Omega}_{s}}}{\partial {t}}.
\label{AA1}
\end{eqnarray}
The substitution of perturbation expansion (\ref{Pert}) in the above equation provides us the force balance at $ O(1) $:
\begin{eqnarray}
&\IB{F}^{(0)} = \int\limits_{{S_p}} {\IB{n} \bcdot {\IB{\sigma}^{(0)} _H}\mathrm{d}S}   +  4\pi H\!a\int\limits_{{S_p}} {\IB{n} \bcdot {\IB{\sigma} _M}\mathrm{d}S}  = \IB{0}\nonumber \\
&\IB{L}^{(0)} = \int\limits_{{S_p}} {\IB{n} \times \left( {\IB{n} \bcdot {\IB{\sigma}^{(0)} _H}} \right)\mathrm{d}S}   +  4\pi H\!a\int\limits_{{S_p}} {\IB{n} \times \left( {\IB{n} \bcdot {\IB{\sigma} _M}} \right)\mathrm{d}S}  = \IB{0},
\label{AA2}
\end{eqnarray}
and at $ O(\Rey_{p}) $:
\begin{eqnarray}
&\IB{F}^{(1)} = \int\limits_{{S_p}} {\IB{n} \bcdot {\IB{\sigma}^{(1)} _H}\mathrm{d}S}   = \frac{4\pi}{3}  \frac{\partial{\IB{U}^{(0)}_{s}}}{\partial {t}}  = \IB{0} \nonumber \\
&\IB{L}^{(1)} = \int\limits_{{S_p}} {\IB{n} \times \left( {\IB{n} \bcdot {\IB{\sigma}^{(1)} _H}} \right)\mathrm{d}S}   = \frac{4\pi}{3} \frac{\partial{\IB{\Omega}^{(0)}_{s}}}{\partial {t}} = \IB{0}.
\label{AA3}
\end{eqnarray}
Since $ \IB{U}_{s}^{(0)} $ and $ \IB{\Omega}_{s}^{(0)} $ are constant, the problems at $ O(1) $ and $ O(\Rey_{p}) $ are force-free and torque free. The translational and angular acceleration of the particle only affect the system at $ O(\Rey_{p}^{2}) $.

%%%%%%%%%%%%%%%%%%%%%%%%%%%%%%%%%%%%%%%%%%%%%%%%%%%%%%%
%%%%%%%%%%%%%%%%%%%%%%%%%%%%%%
%\section{}\label{appB}
\subsection{Illustration of Faxen transformation}
 Since $ \psi_{2} $ depends on $ \psi_{1} $ through the wall boundary condition in (\ref{Pot2}), both reflections must be represented in the coordinates of same scale. The first reflection of potential disturbance in the outer coordinates ($ \tilde{\psi_{1}} $) is:
\begin{equation}
{\tilde \psi _1} =  - \frac{1}{2}\frac{X}{{{R^3}}}{\kappa ^2}.
\label{Psi1Out}
\end{equation}
\citet{faxen1922} represented the fundamental solution of Laplace’s equation (in Cartesian space) in the form of Fourier integrals. Following Faxen, we write:
\begin{equation}
\frac{1}{R} = \frac{1}{{2\pi }}\int\limits_{ - \infty }^{ + \infty } {\int\limits_{ - \infty }^{ + \infty } {{\mathrm{e}^{\left( {\mathrm{i}\Theta  - \frac{{\lambda \left| Z \right|}}{2}} \right)}}\frac{{ \mathrm{d}\xi \mathrm{d}\eta }}{{2\lambda }}} } \mbox{, and\ }
\label{1byR}
\end{equation}
\begin{equation}
R = \frac{1}{{2\pi }}\int\limits_{ - \infty }^{ + \infty } {\int\limits_{ - \infty }^{ + \infty } {\left( {1 - {\mathrm{e}^{\left( {\mathrm{i}\Theta  - \frac{{\lambda \left| Z \right|}}{2}} \right)}}} \right)\left( {1 + \frac{{\lambda \left| Z \right|}}{2}} \right)\frac{{ \mathrm{d}\xi \mathrm{d}\eta }}{{{{{\lambda ^3}} \mathord{\left/
						{\vphantom {{{\lambda ^3}} 2}} \right.
						\kern-\nulldelimiterspace} 2}}}} } .
\label{R}
\end{equation}
Here, $ \Theta  = (\xi X+ \eta Y)/2 $ and $ \lambda=(\xi^{2}+\eta^{2})^{1/2} $. $ \xi $ and $ \eta $ are the variables in Fourier space. Using (\ref{1byR}) and (\ref{R}), we transform the disturbance field (\ref{Psi1Out}) by taking derivatives of the above equations. For example: (\ref{Psi1Out}) contains $ X/R^{3} $, which is expressed using the above transformations as:
\begin{equation}
\frac{X}{{{R^3}}} =  - \frac{\partial }{{\partial X}}\left( {\frac{1}{R}} \right) = \frac{1}{{2\pi }}\int\limits_{ - \infty }^{ + \infty } {\int\limits_{ - \infty }^{ + \infty } {{\mathrm{e}^{\left( {\mathrm{i}\Theta  - \frac{{\lambda \left| Z \right|}}{2}} \right)}}\left( { - \frac{{\mathrm{i}\xi }}{2}} \right)\frac{{ \mathrm{d}\xi \mathrm{d}\eta }}{{2\lambda }}} } .
\label{XbyR3}
\end{equation}
This yields an integral representation for the first reflection:
\begin{equation}
{\tilde \psi _1} = \frac{{{\kappa ^2}}}{{2\pi }}\int\limits_{ - \infty }^{ + \infty } {\int\limits_{ - \infty }^{ + \infty } {{\mathrm{e}^{\left( {\mathrm{i}\Theta  - \frac{{\lambda \left| Z \right|}}{2}} \right)}} {b_1}\left( {\frac{{\mathrm{i}\xi }}{\lambda }} \right) \mathrm{d}\xi \mathrm{d}\eta } }.
\label{Psi1Fax}
\end{equation}
Here, $ b_{1}=1/8 $. The boundary condition expressed in (\ref{Pot2}) suggests a form of solution for $ \tilde{\psi_{2}} $ similar to the above equation:
\begin{equation}
{\tilde \psi _2} = \frac{{{\kappa ^2}}}{{2\pi }}\int\limits_{ - \infty }^{ + \infty } {\int\limits_{ - \infty }^{ + \infty } {{\mathrm{e}^{\mathrm{i}\Theta }}\left( {{\mathrm{e}^{\left( { - \frac{{\lambda Z}}{2}} \right)}}{b_2} + {\mathrm{e}^{\left( { + \frac{{\lambda Z}}{2}} \right)}}{b_3}} \right)\left( {\frac{{\mathrm{i}\xi }}{\lambda }} \right) \mathrm{d}\xi \mathrm{d}\eta } }.
\label{PotSol2App}
\end{equation}
Substituting (\ref{Psi1Fax}) and (\ref{PotSol2App}) in the boundary condition (\ref{Pot2}), yields $ b_{2} $ and $ b_{3} $ in terms of $ b_{1} $:
\begin{equation}
{b_2} = \frac{{{b_1}\left( {1 + {\mathrm{e}^{\left( {1 - s} \right)\lambda }}} \right)}}{{ - 1 + {\mathrm{e}^\lambda }}},\quad {b_3} = \frac{{{b_1}\left( {1 + {\mathrm{e}^{s\lambda }}} \right)}}{{ - 1 + {\mathrm{e}^\lambda }}}.
\label{3.12}
\end{equation}

\subsection{Evaluation of Maxwell stress induced force and torque}
The electric force and torque are expressed as a function of reflections as:
\begin{equation}
{\IB{F}_M} = 4\pi H\!a\int\limits_{{S_p}} {\IB{n} \bcdot \left( \begin{array}{l}
	\bnabla \left( {{\psi _1} + {\psi _2}} \right) \bnabla \left( {{\psi _1} + {\psi _2}} \right)\\
	{\rm{             }} - \frac{1}{2}\left( {\bnabla \left( {{\psi _1} + {\psi _2}} \right) \bcdot \bnabla \left( {{\psi _1} + {\psi _2}} \right)} \right)\mathsfbi I
	\end{array} \right)\mathrm{d}S}, 
\label{FMMor}
\end{equation}
\begin{equation}
{\IB{L}_M} = 4\pi H\!a\int\limits_{{S_p}} {\IB{n} \times \left( {\IB{n} \bcdot \left( \begin{array}{l}
		\bnabla \left( {{\psi _1} + {\psi _2}} \right) \bnabla \left( {{\psi _1} + {\psi _2}} \right)\\
		{\rm{             }} - \frac{1}{2}\left( {\bnabla \left( {{\psi _1} + {\psi _2}} \right) \bcdot \bnabla \left( {{\psi _1} + {\psi _2}} \right)} \right)\mathsfbi I
		\end{array} \right)} \right)\mathrm{d}S}.
\label{LMMor}
\end{equation}
The gradient of first reflection ($ \bnabla \psi_{1} $) at the particle surface is:
\begin{equation}
{\left. {\bnabla {\psi _1}} \right|_{r = 1}} =  \frac{1}{2} \left\{ {-1+3 \sin ^2\theta  \cos ^2\varphi, \; 3 \sin ^2\theta  \sin \varphi  \cos \varphi , \; 3 \sin \theta \cos \theta  \cos \varphi } \right\}.
\label{DelPsi1BC}
\end{equation}
Here, $ \varphi $ and $ \theta $ denote the azimuthal and polar angle, respectively. Since the wall reflected potential ($ \tilde{\psi_{2}} $) has been expressed in the outer coordinates ($ R,X,Y,Z $) in (\ref{PotSol2}), $ \bnabla \psi_{2} $ at the particle surface ($ r=1 $) is expressed as a Taylor series expansion of $ \kappa \tilde{\bnabla} \tilde{\psi_{2}} $ about the origin ($ R \rightarrow 0 $). Here $ \tilde{\bnabla} $ is the spatial gradient in outer scaled coordinates: $ \left\{ {\frac{\partial}{\partial X},\, \frac{\partial}{\partial Y}, \, \frac{\partial}{\partial Z}} \right\} $. Since the particle is small with respect to the wall-particle gap ($ \kappa/s \ll 1 $), we retain only the zeroth and first order terms:
\begin{equation}
{\left. {\bnabla {\psi _2}} \right|_{r = 1}} = \int\limits_0^\infty  {\left[ \begin{array}{l}
	{\kappa ^3}\left( {{{ - ({b_2} + {b_3}){\lambda ^2}} \mathord{\left/
				{\vphantom {{ - ({b_2} + {b_3}){\lambda ^2}} 4}} \right.
				\kern-\nulldelimiterspace} 4}} \right)\left\{ {1, 0, 0} \right\}\\
	+ {\kappa ^4}\left( {{{({b_2} - {b_3}){\lambda ^3}} \mathord{\left/
				{\vphantom {{({b_2} - {b_3}){\lambda ^3}} 8}} \right.
				\kern-\nulldelimiterspace} 8}} \right)\left\{ {1, 0, 1} \right\}
	\end{array} \right]}  \mathrm{d}\lambda   +  O({\kappa ^5}).
\label{DelPsi2BC}
\end{equation}
Substituting (\ref{PotSol1}) $ \& $ (\ref{PotSol2}) in both (\ref{D-F}) $ \& $ (\ref{D-L}) and using (\ref{DelPsi1BC}) $ \& $ (\ref{DelPsi2BC}), we obtain:
\begin{equation}
{\IB{F}_M} = 4\pi H\!a\left( {\frac{{3\pi }}{{16}}{\kappa ^4}\left(\mathfrak{Z}{\left({4,s} \right) - \mathfrak{Z} \left( {4,1 - s} \right)} \right)} \right) \IB{e}_{z}  +  O\left( {{\kappa ^7}} \right)\; \text{and} \; {\IB{L}_M} = \IB{0}.
\label{FMLMApp}
\end{equation}
The leading order electrical force on the particle arises only due to the interaction of $ O(1) $ source-dipole term ($ \bnabla \psi_{1} $) and $ O(\kappa^{4}) $ term in ($ \bnabla \psi_{2} $). This was also reported by \citet{yariv2006} for the case of a particle remotely adjacent to a single wall (i.e. $ \kappa \ll s $).

\subsection{Solution to $\IB{v}_{3}^{(0)}$}
Here, we provide the details of the evaluation of $ \IB{v}_{3}^{(0)} $ (see \S\ref{sec:3}.3).
The particle boundary condition in (\ref{Vel3}) requires us to calculate $ \tilde{\IB{v}}_{2}^{(0)} $ in the vicinity of the particle. Since, $ \tilde{\IB{v}}_{2}^{(0)} $ is represented in outer scaled coordinates, the particle surface is equivalent to $ R \rightarrow 0 $. Upon expanding $ \tilde{\IB{v}}_{2}^{(0)} $ about the origin, we obtain:
\begin{equation}
{\left. {\IB{v}_2^{\left( 0 \right)}} \right|_{r = 1}} =  \int\limits_0^{\infty } {\left[ \begin{array}{l}
	\left({2{\ell _4} + {\ell _5} + 2{\ell _7} + {\ell _8}} \right)\lambda/2 \; - \; \left( {2{\ell _4} + {\ell _5} - {\ell _6} - 2{\ell _7} - {\ell _8} + {\ell _9}} \right){\lambda ^2}\kappa  z/4 \\ [2 pt]
	\quad {\rm{       }} + {\kappa ^3}Ha{Z_w}\, \sl{2} + \cdots \\ [6 pt]
	0 + \cdots\\ [6 pt]
	\left( { - {\ell _4} - {\ell _5} - {\ell _6} + {\ell _7} + {\ell _8} + {\ell _9}} \right){\lambda ^2}\kappa  x/4 + \cdots
	\end{array} \right]} \mathrm{d}{\theta _C}
\label{A4}
\end{equation}
The derivation of $ \tilde{\IB{v}}_{2}^{(0)} $ along with the evaluation $ \ell_4,\: \ell_5, \cdots \ell_9 $ are described in supplementary material. Having obtained the boundary condition for $ \IB{v}_{3}^{(0)} $ at the particle surface, we use Lamb's method to obtain the third reflection. The resulting solution has a form similar to (\ref{VelSol1}) with coefficients $ A_{3},\:B_{3}, \:C_{3}\mbox{ and\ } D_{3} $. These coefficients are in the integral form, owing to the integral form of (\ref{A4}). Since the force-free and torque free arguments require only the coefficients of stokeslet and rotlet field \citep[p. 88]{kim2013}, here we only report the coefficients $ A_{3} $ and $ C_{3} $ for brevity:
\begin{equation}
\begin{array}{l}
{A_3} = \frac{{ - 3}}{4}\int\limits_0^\infty  {\left( {\frac{1}{2}\left( {2{\ell _4} + {\ell _5} + 2{\ell _7} + {\ell _8}} \right)\lambda  + {\kappa ^3}Ha{{\rm Z}_w}{_x}\sl{2}} \right)} \mathrm{d}\lambda \\ [3 pt]
{C_3} = \frac{1}{4}\int\limits_0^\infty  {\left( {\frac{1}{2}\left( {{\ell _4} - 2{\ell _6} - {\ell _7} + 2{\ell _9}} \right){\lambda ^2}\kappa } \right)} \mathrm{d}\lambda 
\end{array}
\label{A5}
\end{equation}
Substitution of $ \ell_4,\: \ell_5, \cdots \ell_9 $ into the above equations results in (\ref{A3})-(\ref{C3}).

\subsection{Test field ($\IB{u}^{t}$)}
Using the method of reflections we seek:
\begin{equation}
{\IB{u}^t} = \IB{u}_1^t + \IB{u}_2^t + \IB{u}_3^t +\cdots \qquad {p^t} = p_1^t + p_2^t + p_3^t + \cdots
\label{A6}
\end{equation}
Following \citet{ho1974}, we obtain the first reflection $ \IB{u}^{t}_{1} $:
\begin{equation}
\IB{u}_1^t = \frac{3}{4}\left( {{\IB{e}_z} + \frac{{z\IB{r}}}{{{r^2}}}} \right)\frac{1}{r} + \frac{1}{4}\left( {{\IB{e}_z} - \frac{{3z\IB{r}}}{{{r^2}}}} \right)\frac{1}{{{r^3}}}
\label{A7}
\end{equation}
and the second reflection $ \tilde{\IB{u}}^{t}_{1}=\{\tilde{u}^{t}_{1},\:\tilde{v}^{t}_{1},\:\tilde{w}^{t}_{1}\} $:
\begin{equation}
\tilde u_2^t = \frac{1}{{2\pi }}\int_{ - \infty }^{ + \infty } {\int_{ - \infty }^{ + \infty } {{e^{\rm{i}{\Theta ^t}}}\left( {{e^{\left( {\frac{{ - {\lambda ^t}Z}}{2}} \right)}}\left( {\frac{{{h_3}Z}}{2} + \frac{{{h_4}}}{{{\lambda ^t}}}} \right) + {e^{\left( {\frac{{ + {\lambda ^t}Z}}{2}} \right)}}\left( {\frac{{{h_5}Z}}{2} - \frac{{{h_6}}}{{{\lambda ^t}}}} \right)} \right)\rm{i}{\xi ^t} d{\xi ^t}} } d{\eta ^t}
\label{A8}
\end{equation}
\begin{equation}
\tilde v_2^t = \frac{1}{{2\pi }}\int_{ - \infty }^{ + \infty } {\int_{ - \infty }^{ + \infty } {{e^{i{\Theta ^t}}}\left( {{e^{\left( {\frac{{ - {\lambda ^t}Z}}{2}} \right)}}\left( {\frac{{{h_3}Z}}{2} + \frac{{{h_4}}}{{{\lambda ^t}}}} \right) + {e^{\left( {\frac{{ + {\lambda ^t}Z}}{2}} \right)}}\left( {\frac{{{h_5}Z}}{2} - \frac{{{h_6}}}{{{\lambda ^t}}}} \right)} \right)i{\eta ^t} d{\xi ^t}} } d{\eta ^t}
\label{A9}
\end{equation}
\begin{equation}
\tilde w_2^t = \frac{1}{{2\pi }}\int_{ - \infty }^{ + \infty } {\int_{ - \infty }^{ + \infty } {{e^{i{\Theta ^t}}}\left( \begin{array}{l}
		- {e^{\left( {\frac{{ - {\lambda ^t}Z}}{2}} \right)}}\left( {{h_3}\left( {1 + {{{\lambda ^t}Z} \mathord{\left/
						{\vphantom {{{\lambda ^t}Z} 2}} \right.
						\kern-\nulldelimiterspace} 2}} \right) + {h_4}} \right)\\
		- {e^{\left( {\frac{{ + {\lambda ^t}Z}}{2}} \right)}}\left( {{h_5}\left( {1 - {{{\lambda ^t}Z} \mathord{\left/
						{\vphantom {{{\lambda ^t}Z} 2}} \right.
						\kern-\nulldelimiterspace} 2}} \right) + {h_6}} \right)
		\end{array} \right)d{\xi ^t}} } d{\eta ^t}
\label{A10}
\end{equation}
Here, $ \Theta=(X\xi^{t}+Y\eta^{t})/2 $ and $ \lambda^{t}=(\xi^{t \, ^2}+\eta^{t \, ^2})^{1/2} $. The expressions for the terms $ h_{3},\: h_{4},\cdots h_{6} $ are evaluated and are found to be identical to that reported by \citet{ho1974}. The disturbance velocities can be reduced to single integral by the use of simple transformation techniques, which makes the evaluation inertial lift convenient. An interested reader can refer the supplementary material.

%%%%%%%%%%%%%%%%%%%%%%%%%%%%%%%%%%%%%%%%%%%%%%%%%%%%%%%
%%%%%%%%%%%%%%%%%%%%%%%%%%%%%%
\section{Validation of $ \IB{U}_{s} $ and $ \IB{\Omega}_{s} $ with the literature}\label{appB}
In the absence of electric fields, the expressions for $ U_{s\,x}^{(0)} $ (\ref{U0}) and $ \Omega_{s\,y}^{(0)} $ (\ref{Om0}) are identical to that obtained by \citet{ho1974}. 

For the pure electrokinetic flow (i.e. electrophoretic particle in electro-osmotic flow: $ \alpha=0,\: \beta=0,\: \mbox{and\ }\gamma=0 $), the dimensional electrophoretic translational velocity ($ \IB{U}_{s}^{'\,E} $) at the centerline ($ s=0.5 $) is: 
\begin{equation}
\frac{\IB{U}_{s}^{'E}}{E_{\infty}'(\varepsilon'/4\pi \mu')} = [ \,  (\zeta_p' - \zeta_w') - (0.259478 \zeta_p' - 0.262433 \zeta_w')(a'/d')^{3} \, ] \IB{e}_{x}.
\label{UE0}
\end{equation}
For a particle fixed at the centerline the expression derived by \citet{keh1985} in bounded electrokinetic flow is: 
\begin{equation}
\frac{\IB{U}_{s}^{'E}}{E_{\infty}'(\varepsilon'/4\pi \mu')} = [ \, 1-0.267699 (a'/d')^{3} + O(a'/d')^{5} \, ]  (\zeta_p' - \zeta_w') \IB{e}_{x}.
\label{UE0Keh}
\end{equation}
This difference arises as \citet{keh1985} use an alternative method of reflections whereas, we follow Brenner’s parallel method of reflections (see \citep{luke1989} for more details on the two methods). When the resistance due to stokeslet ($ W_{A} $) is neglected in (\ref{U0}), for pure electrokinetic flow at $ s=0.5 $, (\ref{UE0Keh}) is obtained. Since (\ref{U0})---for pure electrokinetic flow---is valid for any arbitrary position between the walls, it is a generalization of results obtained by \citet{keh1985} upto $ O(\kappa^{3}) $.
\begin{figure}
	\centerline{\includegraphics[scale=0.3]{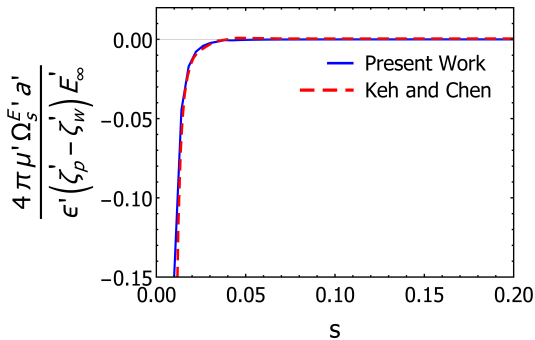}}% Images in 100% size
	\caption{Variation of the electrophoretic rotational velocity with respect to wall separation for \textit{a}=10 $\mu$m.}
	\label{fig:3}
\end{figure}

For the pure electrokinetic flow, the weak particle rotation ($ \IB{\Omega}_{s}^{E} \sim O(\kappa^{4}) $) near the walls is negative as depicted by figure \ref{fig:3}. This is interestingly opposite to that observed in Couette, Poiseuille flow and sedimentation flows. Near the wall, the potential distribution around the particle exerts a hydrodynamic torque (through slip at the particle surface, see (\ref{SlipCorr})) in the negative $ y $-direction, and is sufficiently strong to overcome  torque generated due to wall induced viscous resistance \citep{keh1988}. We compare the rotational velocity expressed in (\ref{Om0}) with that obtained \textit{exactly} by \citet{keh1988} (in the presence of a single wall) and find a good agreement (see figure \ref{fig:3}).
\vspace{5 mm}

%%%%%%%%%%%%%%%%%%%%%%%%%%%%%
%%%%%%%%%%%%%%%%
\section{}\label{appD}
In this appendix, we analytically find the contribution from the sub-region which extends from the spherical surface to the circumscribing cylinder (see \S \ref{sec:4}.3). The spherical coordinate system is employed to find the volume integral as shown in figure \ref{fig:11}(a).
The azimuthal angle $ \rm{\phi}_s $ spans from 0 to 2$ \pi $. The polar angle $ \psi_{s} $ (spanning from 0 to $ \pi $) is divided into 4 sections. This splitting results in four right angled triangles and allows us to exploit trigonometric identities to specify limits in the radial direction $ \left( {\int\limits_\kappa ^{{r_{out}}} {dr} } \right) $. For instance, in the section-I (0 to $ \pi $/4), we obtain the following relation using trigonometry: $ \cos \psi_{s}=\kappa/r_{out} $. Therefore, the radius of integration spans from $ \kappa \to \kappa \sec \psi_{s} $. Extending the procedure to all the four sections, transforms the volume integral as: 

\begin{figure}
	\centerline{\includegraphics[scale=0.42]{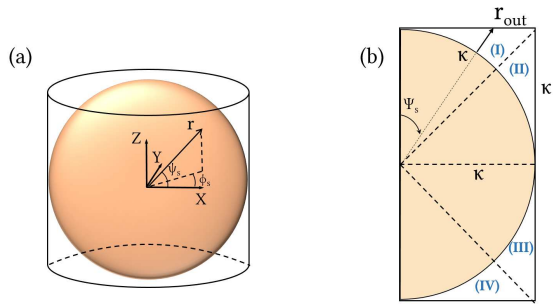}}% Images in 100% size
	\caption{(a) Cylinder circumscribing the spherical particle of radius $ \kappa $, (b) front 2D projection showing the splitting of polar angle into 4 identical sections. The integral is evaluated in the blank space. Here, $ r_{out} $ is the outer limit of radius.}
	\label{fig:11}
\end{figure}

\begin{equation}
\epsilon = {\int_{ 0 }^{ 2 \pi } {\left( \begin{array}{l}
		\left( {\mathop \smallint \limits_0^{\pi /4}  \mathop \smallint \limits_\kappa^{\kappa \sec \psi_{s} } {\mathcal{H}}\left( {{r^2}\sin \psi_{s} } \right)drd\psi_{s} } \right) + \left( {\mathop \smallint \limits_{\pi /4}^{\pi /2}  \mathop \smallint \limits_\kappa^{\kappa \csc \psi_{s} } {\mathcal{H}}\left( {{r^2}\sin \psi_{s} } \right)drd\psi_{s} } \right)\\
		+ \left( {\mathop \smallint \limits_{\pi /2}^{3\pi /4}  \mathop \smallint \limits_\kappa^{\kappa \csc \psi_{s} } {\mathcal{H}}\left( {{r^2}\sin \psi_{s} } \right)drd\psi_{s} } \right) + \left( {\mathop \smallint \limits_{3\pi /4}^\pi  \mathop \smallint \limits_\kappa^{ - \kappa \sec \psi_{s} } {\mathcal{H}}\left( {{r^2}\sin \psi_{s} } \right)drd\psi_{s} } \right)
		\end{array} \right)d\rm{\phi}_s } } \\[2 pt]
\label{E1}
\end{equation}
Here, $ {\mathcal{H}} $ is the integrand in spherical coordinates. The contribution from the four sections is found to be:
\begin{equation}
\epsilon  =  - \Rey_{p}\beta \left( {1.7 \,Ha{Z_p} + 1.603\kappa }+O(\kappa^{2}) \right),
\label{E2}
\end{equation}
and is accounted in the results discussed in \S \ref{sec:7}. For the case of neutrally and non-neutrally buoyant (inert) particle, we find that the contribution from this sub-region is negligible.

\bibliographystyle{jfm}
\bibliography{Akash}

\end{document}